\PassOptionsToPackage{dvipsnames}{xcolor}
\PassOptionsToPackage{a4paper, bottom=0cm}{geometry}
\documentclass[prx,12pt,a4paper,twocolumn,margin=0.1cm,bottom=0.1cm,unpublished,accepted=2026-08-03]{quantumarticle}
\pdfoutput=1
\usepackage[sort&compress,numbers]{natbib}
\usepackage[T1]{fontenc}
\usepackage[utf8]{inputenx}

\usepackage{amsmath,amssymb,amsthm,bm,amsfonts,mathrsfs,bbm}
\usepackage{thm-restate}
\usepackage{xspace}
\usepackage{changes}
\usepackage{minted} 
\usepackage{algorithm}
\usepackage[noend]{algpseudocode}
\usepackage{braket} 
\usepackage{wrapfig}
\usepackage[english]{babel} 
\usepackage[autostyle=true, autopunct=true]{csquotes}
\usepackage{xcolor}
\usepackage[colorlinks = true,
            linkcolor = blue,
            urlcolor  = blue,
            citecolor = blue,
            anchorcolor = blue]{hyperref} 
\usepackage{graphicx}
\usepackage[babel]{microtype} 
\microtypecontext{spacing=nonfrench}
\microtypesetup{
expansion={true,nocompatibility},
protrusion={true,nocompatibility},
activate={true,nocompatibility},
tracking=false,
kerning=true,
spacing={true}
}

\usepackage[all=normal,floats=tight,mathspacing=tight,wordspacing=tight,paragraphs=normal,tracking=tight,charwidths=tight,mathdisplays=normal,sections=normal,margins=normal]{savetrees}

\usepackage{newtxtext}

\usepackage[nottoc]{tocbibind} 
\usepackage{subcaption}
\usepackage{mathtools}
\usepackage{centernot}
\usepackage[arrowdel]{physics}
\usepackage{bbold}
\usepackage{todonotes}
\usepackage{comment}
\usepackage{paralist}

\makeatletter
\def\thmheadbrackets#1#2#3{%
  \thmname{#1}\thmnumber{\@ifnotempty{#1}{ }\@upn{#2}}%
  \thmnote{ {\the\thm@notefont[#3]}}}
\makeatother
\newtheoremstyle{definition}
  {}
  {}
  {}
  {}
  {}
  {\textrm{:}}
  { }
  {\thmheadbrackets{\textbf{#1}}{#2}{\bfseries \textit{#3}}}
\theoremstyle{definition}

\newtheorem*{definition}{Definition}
\newtheorem*{informaldefinition}{Informal Definition}

\newtheoremstyle{theorem}
  {}
  {}
  {}
  {}
  {\bfseries}
  {:}
  { }
  {\thmheadbrackets{\textbf{#1}}{#2}{\textbf{#3}}}
\theoremstyle{theorem}
\newtheorem{prop}{Proposition}
\newtheorem{theorem}{Theorem}

\definecolor{ao(english)}{rgb}{0.0, 0.5, 0.0}
\newcommand{\cris}[1]{{\color{ao(english)} C: #1}}
\newcommand{\maria}[1]{{\color{violet} #1}}
\newcommand{\paolo}[1]{{\color{blue} #1}}

\usepackage{xspace}
\newcommand{\mnn}{\textup{\textsf{M\!N\!N}}\xspace}
\newcommand{\Shadowed}{\textup{\textsf{S\kern-0.1em h\kern-0.1em a\kern-0.1em d\kern-0.1em o\kern-0.1em w\kern-0.1em e\kern-0.1em d}}\xspace}
\newcommand{\Shadow}{\textup{\textsf{S\kern-0.1em h\kern-0.1em a\kern-0.1em d\kern-0.1em o\kern-0.1em w}}\xspace}
\newcommand{\SUN}{\textup{\textsf{S\!U\!N}}\xspace}
\let\MNN\mnn
\let\UNCLE\SUN
\let\CABW\Shadowed
\newcommand{\fnn}{\textup{\textsf{F\!N\!N}}\xspace}

\title{Escaping the Shadow of Bell's Theorem in Network Nonlocality}
\author{Maria Ciudad Alañón}
\email{mciudadalanon@pitp.ca}
\affiliation{Perimeter Institute for Theoretical Physics, 31 Caroline Street North, Waterloo, Ontario, Canada N2L 2Y5}
\affiliation{Department of Physics and Astronomy, University of Waterloo, Waterloo, Ontario, Canada, N2L 3G1}
\author{Emanuel-Cristian Boghiu}
\email{cristian.boghiu@icfo.eu}
\affiliation{ICFO – Institut de Ciencies Fotoniques, The Barcelona Institute of Science and Technology, 08860 Castelldefels (Barcelona), Spain}
\author{Paolo Abiuso}
\email{paolo.abiuso@oeaw.ac.at}
\affiliation{Institute for Quantum Optics and Quantum Information - IQOQI Vienna, Austrian Academy of Sciences, Boltzmanngasse 3, A-1090 Vienna, Austria}
\author{Elie Wolfe}
\email{ewolfe@pitp.ca}
\affiliation{Perimeter Institute for Theoretical Physics, 31 Caroline Street North, Waterloo, Ontario, Canada N2L 2Y5}
\affiliation{Department of Physics and Astronomy, University of Waterloo, Waterloo, Ontario, Canada, N2L 3G1}

\begin{document}

\maketitle

\begin{abstract}
The possibility of nonclassicality in networks unrelated to Bell’s original eponymous theorem has recently attracted significant interest. Here, we identify a sufficient condition for being “outside the shadow of Bell’s theorem” and introduce a testable criterion capable of certifying the novelty of instances of network-nonclassicality which we call minimal network nonclassicality. We provide examples of minimally network nonclassical correlations realizable in quantum theory as well as examples coming from more exotic operational probabilistic theories. In particular, we apply these concepts to the simplest configuration of the 3-chain scenario  (a.k.a. the bilocality scenario) to prove that certain correlations have escaped the shadow of Bell's theorem. While some of the examples herein are unprecedented, we also revisit more familiar examples of network nonclassicality in order to highlight the contrast between our approach versus prior approaches with respect to assessing novelty.

\end{abstract}

\section{Introduction}
In 1964, John Bell showed that quantum theory leads to correlations that cannot be explained by any theory that satisfies the assumptions of measurement independence and local causality~\cite{bell1964einstein, wiseman2017causarum}. This phenomenon --- which we call \textit{Bell-type nonclassicality}\footnote{Also known as Bell nonlocality.} --- led to the development of an entire field of research aimed at understanding and resolving the conflict between these assumptions and quantum theory. Considerable progress has been made in this area, particularly in understanding how various nonclassical features of quantum theory manifest as Bell-type nonclassicality, and in showing how this nonclassicality can become a resource for quantum information processing tasks, such as randomness generation or physics-based cryptography (see Ref.\,~\cite{RevModPhys.86.419} for a review).

\begin{figure}[h]
\centering
\includegraphics[width=0.4\columnwidth]{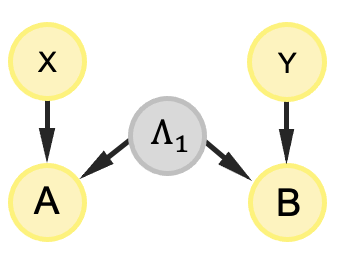}
\caption{Representation of the standard Bell's scenario, where $A, B$ represent observed variables, $X, Y$ locally generated settings and $\Lambda$, a latent variable. The latent variables could be either a classical variable, a quantum system or a more general source of correlation.}
\label{fig:intro:bell}
\end{figure}

The original scenario considered by Bell consists of two separated parties each locally performing independently chosen measurements in their laboratories. As the two laboratories cannot communicate instantly as per the finite speed of light, correlations among their outcomes can only arise from previously shared bipartite sources (including e.g., entangled quantum states). When restricting to classical theory, these assumptions of measurement independence and local causality are equivalent to the existence of a local hidden-variable description of the experiment, i.e., a classical common-cause explanation of the observed measurement correlations (Fig.~\ref{fig:intro:bell}). Bell nonclassicality, then, is the property of certain experiments featuring quantum nonseparable sources to result in outcomes inexplicable by any such classical common-cause model.
The same situation can be immediately extended to the case of $N>2$ isolated parties sharing $1$ common source, which we shall include here in the terminology of \emph{standard Bell scenario}.

In recent years it was observed, however, that the single common-cause explanation of an $N$-partite experiment, is a specific example within a broader spectrum of \textit{causal models}, which can be systematically studied with the tools provided by the field of causal inference~\cite{pearl2009causality}. Causal models consist of observable random variables with known probability distributions and latent variables with unknown distributions which, together, explain observed correlations through causal influences, i.e., functional relationships between the values of different variables. 

Causal models have gained attention in the field of quantum information theory as they provide a natural generalisation of Bell's scenario to other scenarios of interest, such as those of \textit{quantum networks} of space-like separated parties sharing independent quantum states~\cite{tavakoli2022bell}. These networks allow for novel quantum phenomena that have no analogue in Bell's scenario, such as the possibility of performing entanglement swapping, which can distribute entanglement between distant, initially independent parties in the network~\cite{PhysRevLett.71.4287}.  

Thus, causal models have been generalised to include quantum sources of correlations~\cite{Allen_2017, barrett2020quantum}. Contrasting the correlations achievable in a causal model with either classical or quantum sources allows for the identification of \textit{quantum advantage} in a given network topology, understood as the generation of network correlations that cannot be realised with classical resources alone. Causal models have also been generalised to allow for sources of correlations from other theories, possibly beyond quantum mechanics, such as theories constrained only by no-superluminal signalling~\cite{gisin2020constraints,coiteux-roy2021any}. 
The causal network approach to quantum (and post quantum) nonclassicality has proven to be powerful in the analysis of foundational questions~\cite{tavakoli2022bell,renou2021quantum,coiteux-roy2021any,Weilenmann_2020, Weilenmann_2020_2}, and it provides new tools for understanding information processing in a theory-agnostic way.

The study of causal models with quantum resources naturally suggest certain open questions.
It is in particular natural to ask:
\begin{compactenum}
\item What are the simplest scenarios of nonclassicality? 
\item When is it possible to guarantee a network nonclassical behaviour is also nontrivial, i.e., not merely a logical consequence of quantum advantages in standard Bell scenarios?
\end{compactenum}
The latter question can be rephrased as follows: Given some observed nonclassical correlations in a network, can we confidently distinguish whether their nonclassicality is ultimately traceable to Bell-type nonclassicality, or whether it represents a \textit{new type of network nonclassicality}? 
We here, as also noted by others 
\cite{_upi__2022,PhysRevLett.130.190201,renou2019genuine}, highlight that many significant correlations in networks are such that their nonclassicality \emph{can} be traced back to standard Bell nonclassicality (cf. Sec.~\ref{sec:beyond_bell} for details). 

These questions have led to various studies and proposed definitions on what one should consider as \emph{genuine} network nonclassicality. The approach of~\citet{_upi__2022} is to define non-genuine nonclassical quantum correlations as those realisable in a network where the parties perform local measurements on the sources they receive, and local wirings of their local measurement settings and outcomes. Some examples of correlations have been proven to be outside this set (i.e. genuine network nonclassical according to Ref.~\cite{_upi__2022}), particularly by certifying, through self-testing~\cite{_upi__2020}, the use of entangling measurements~\cite{PhysRevLett.121.250507, PhysRevLett.121.250506} or the use of non-entangling measurements that cannot be implemented with local operations and classical communication~\cite{_upi__2023}. Nevertheless, while self-testing represents the highest standard of nonclassicality benchmark in quantum networks, existing techniques based on it are technically challenging and can only detect very pure sources of entanglement. 
Moreover, we shall show that certain correlations arising \emph{only via entangled measurements} can nevertheless be seen as simple liftings of standard Bell violations.

A different approach can be found in Refs.~\cite{renou2019genuine, pozas2023proofs, abiuso2022single}.  They aim to depart from the standard Bell scenario by identifying nonclassical correlations manifesting in networks without independent measurement choices. 
However, the absence of measurement choices is insufficient to establish Bell-unrelatedness. \citet[Thm. 2.16]{Fritz_2012} shows how Bell-type nonclassicality can be embedded in a larger network to produce nonclassicality without measurement choices. 

Yet another approach is given by~\citet{Pozas_Kerstjens_2022}, who 
propose a theory-independent definition of \textit{full network nonclassicality}\footnote{The authors originally named it \textit{fully network nonlocality}. We have elected to substitute the word \enquote{nonlocality} for \enquote{nonclassicality}, as \enquote{nonlocality} carries some connotation of unhelpful philosophical baggage.} (\fnn),  describing the set of correlations that requires \textit{all} sources in the network to be nonclassical in order to manifest. 
Full network nonclassicality can be certified by linear programming tests based on the inflation technique~\cite{Wolfe_2019, Navascues_2020}. In addition, the tests are noise robust, facilitating the experimental demonstration~\cite{PhysRevLett.130.190201}. However, we will argue that \fnn includes certain correlations which we consider as traceable to the standard Bell scenario, demonstrating that full network nonclassicality is therefore \emph{not} a proxy for Bell-unrelatedness.

In this work we also take up the challenge of articulating what makes a network-nonclassical correlation truly novel versus nonclassicality \enquote{in the shadow of Bell's theorem}. We will not claim to provide a single precise definition here --that task is left for future work-- but the concept, however vague, is worth naming. We say that a correlation \enquote{lives in the shadow of Bell} (hereafter referred to as \CABW) whenever the nonclassicality of the correlation can, in one or more ways, be traced back to Bell’s original theorem regarding the extraordinary explanatory power of a single fixed nonclassical common cause.
That is, any \emph{proof} of nonclassicality that views the network's correlation from the perspective of a traditional proof of Bell's theorem in the standard Bell scenario casts a shadow on the correlation. A correlation is said to be \emph{in more than one shadow} if its nonclassicality can be seen from multiple different angles via Bell's theorem, that is, if there is more than one way to prove the nonclassicality via Bell's theorem.

Here we do not attempt to precisely delineate the boundary between novel-or-not network nonclassicality. Rather, we identify and formalize a common feature shared by all \CABW correlations. This, in turn, suggests a specialized assessment for the purpose of irrefutably witnessing the novelty (non-{\CABW}ness) of a given correlation. We deem this novelty-certifying property \emph{Minimal Network Nonclassicality} (\mnn). It is defined in a theory-agnostic manner, consistent with that of Ref.~\cite{Pozas_Kerstjens_2022}.
Our definition is constructed so as to ensure that \mnn\ correlation cannot be simulated via underlying embeddings of standard Bell nonclassicality. Indeed, our definition of \MNN will be motivated by first appreciating that all \CABW correlations --- read expansively --- share a common feature. The \emph{absence} of this feature in all \MNN correlations within networks containing only bipartite nonclassical sources is what makes our definition of \MNN valuable for certifying genuinely novel instances of network nonclassicality.

\section{Beyond Bell nonclassicality}
\label{sec:beyond_bell}

Our starting observation consists in noticing that, among
several known examples of nonclassicality in  networks,
many can be seen to leverage simpler Bell-type nonclassicality cleverly embedded into a non-Bell-scenario network.  
This class of strategies is what we have termed \CABW. A notable example of a \CABW embedding was pointed out by \citet{Fritz_2012}, showing how two parties violating the Clauser-Horne-Shimony-Holt (CHSH) inequality can be classically wired to a third party to violate locality in the resulting triangle network.

\begin{figure}
        \begin{subfigure}[b]{0.25\textwidth}
                \centering
                \includegraphics[width=0.8\linewidth]{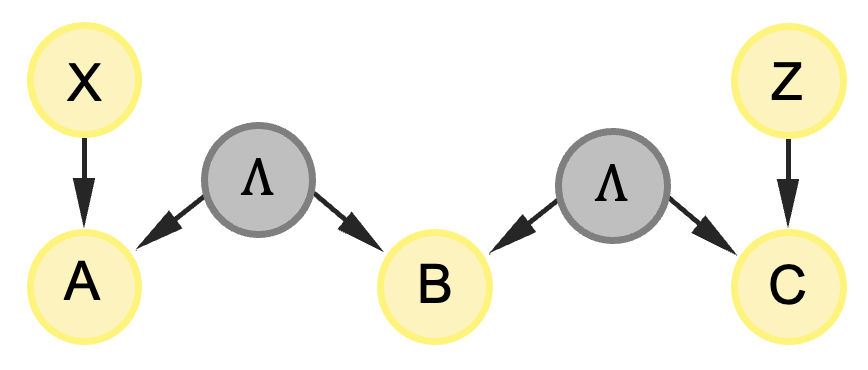}
                \caption{3-chain scenario}
                \label{fig:ex_bilocality}
        \end{subfigure}%
        \begin{subfigure}[b]{0.25\textwidth}
                \centering
                \includegraphics[width=0.8\linewidth]{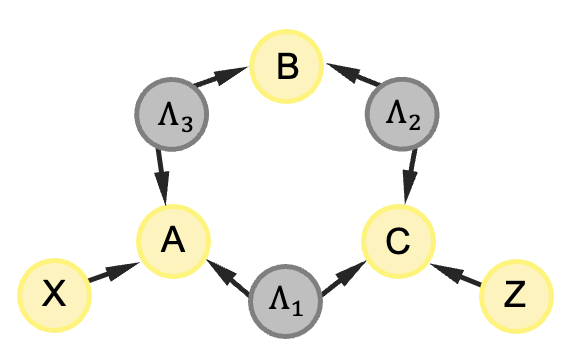}
                \caption{Triangle scenario}
                \label{fig:ex_triangle}
        \end{subfigure}%
        \caption{Representation of the causal structure of the 3-chain scenario where $B$ has no inputs and the triangle scenario where $A$ and $C$ have inputs.}
\label{fig:examples} 
\end{figure}

Moreover, in the scenario depicted in Fig.~\ref{fig:ex_bilocality}, which we refer to as \emph{3-chain scenario} (also known as the  \emph{bilocality} scenario~\cite{branciard2010characterizing,branciard2012bilocal,tavakoli2022bell}), entanglement-swapping 
can be exerted to exhibit nonclassicality by means of (e.g.) CHSH violations between the extremal parties when post-selecting on successful teleportation by the middle party. This too, in our view, is an example of nonclassicality which can be traced back to Bell's original theorem.

A further insightful example consists of the triangle scenario where two (say $A$ and $C$) out of the three parties have settings~\ref{fig:ex_triangle} and the three parties are performing the entanglement-swapping protocol (details in Appendix~\ref{app:shadow_examples}). In this case, the nonclassicality of the correlation between $A$ and $C$ can be explained via two alternative ways: either through the nonclassicality of the source between $A$ and $C$ or via the post-selection of the party $B$ between them. Despite the multiplicity of explanations for the same correlation in this scenario, \emph{all} the explanations appear to trace back to Bell's theorem at some level.

We detail these examples, together with the inequalities that detect their nonclassicality in Appendix~\ref{app:shadow_examples}. It is therefore natural to ponder whether novel forms of nonclassicality arise from the structure of the network rather than being attributed to a Bell-nonclassicality embedding.
In the above-mentioned examples (as detailed in App.~\ref{app:shadow_examples}), 
all the \CABW explanations of the violation of classicality can be seen to ultimately leverage some \emph{fixed} effective nonclassical source(s), either part(s) of the original network or induced by measurements. For instance, the entanglement-swapping example in the 3-chain network exploits entanglement between the extremal parties induced by post-selection on (i.e. conditioning on a particular outcome of) an entangled measurement. 

Note that all the \CABW explanations rely on (at least) one fixed \emph{effective} nonclassical common cause, but not all of them hinge of the nonclassicality of the same physical common cause (as is the case in the last example mentioned). Indeed, a fuller defense of interpreting Bell´s theorem as witnessing the nonclassicality of a common cause can be found in the introduction of Ref.~\cite{wolfe2020quantifying}.

In other words, all the \CABW explanations of the nonclassical correlation between two parties rely on finding a \emph{nonclassical} path between those two parties. Such nonclassical paths can be present in the original structure of the network or it can appear by performing a measurement (specifically, by conditioning on a particular outcome of a measurement). More details about the notion of nonclassical paths can be found in Appendix~\ref{app: nonclassical_paths}

We use this observation to define a set of nonclassical distributions which include all (but perhaps not \emph{only}) \CABW strategies: 

\begin{informaldefinition}[\textsf{S}ubset \textsf{U}navoidable \textsf{N}onclassicality (\UNCLE)]
A correlation is said to exhibit \textsf{S}ubset \textsf{U}navoidable \textsf{N}onclassicality (herafter \UNCLE) if all explanations for realizing it in the given network --- allowing for nonclassical resources --- rely on some fixed subset of the parties exploiting a nonclassical common cause, where we do not distinguish between a nonclassical common cause which is originally present in the network as a source from one which is induced via post-selection. See Appendix~\ref{app: nonclassical_paths} for the \emph{formal} definition of \UNCLE in terms of nonclassical paths.
\end{informaldefinition}
That is, a correlation exhibits \UNCLE when it \emph{cannot} be explained \emph{without} particular parties making use of some originally-present or postselection-induced nonclassical common cause. 
Notice that, for a given correlation, being \UNCLE is network dependent. As with \CABW, a correlation may exhibit \emph{multiple} instances of \UNCLE if a fixed \emph{set} of common causes must be nonclassical in order to explain the correlation.

\begin{prop}\label{prop:CABWimpliesUNCLE}
If a network nonclassical correlation can be traced back to Bell's theorem in the standard Bell scenario, then it certainly must be such that all explanations for the correlation rely on some particular subset of the parties making use of a nonclassical common cause (either an originally-present nonclassical source or a nonclassical common cause induced by postselection). 
\\\indent\begin{minipage}[t]{0.93\linewidth}
\indent\textbf{Shorthand:} ${\CABW \implies \UNCLE}$
\\\indent\textbf{Mnemonic:} To be in \Shadow the \UNCLE must be present.\footref{disclaimer}
\end{minipage}\end{prop}\footnotetext{\label{disclaimer}All mnemonics are intended as memory aids only. Critical readers will surely note that the claims are not actually true relative to the plain (nontechnical) meanings of the words.}

\begin{figure}[t]
\centering
\includegraphics[width=\columnwidth]{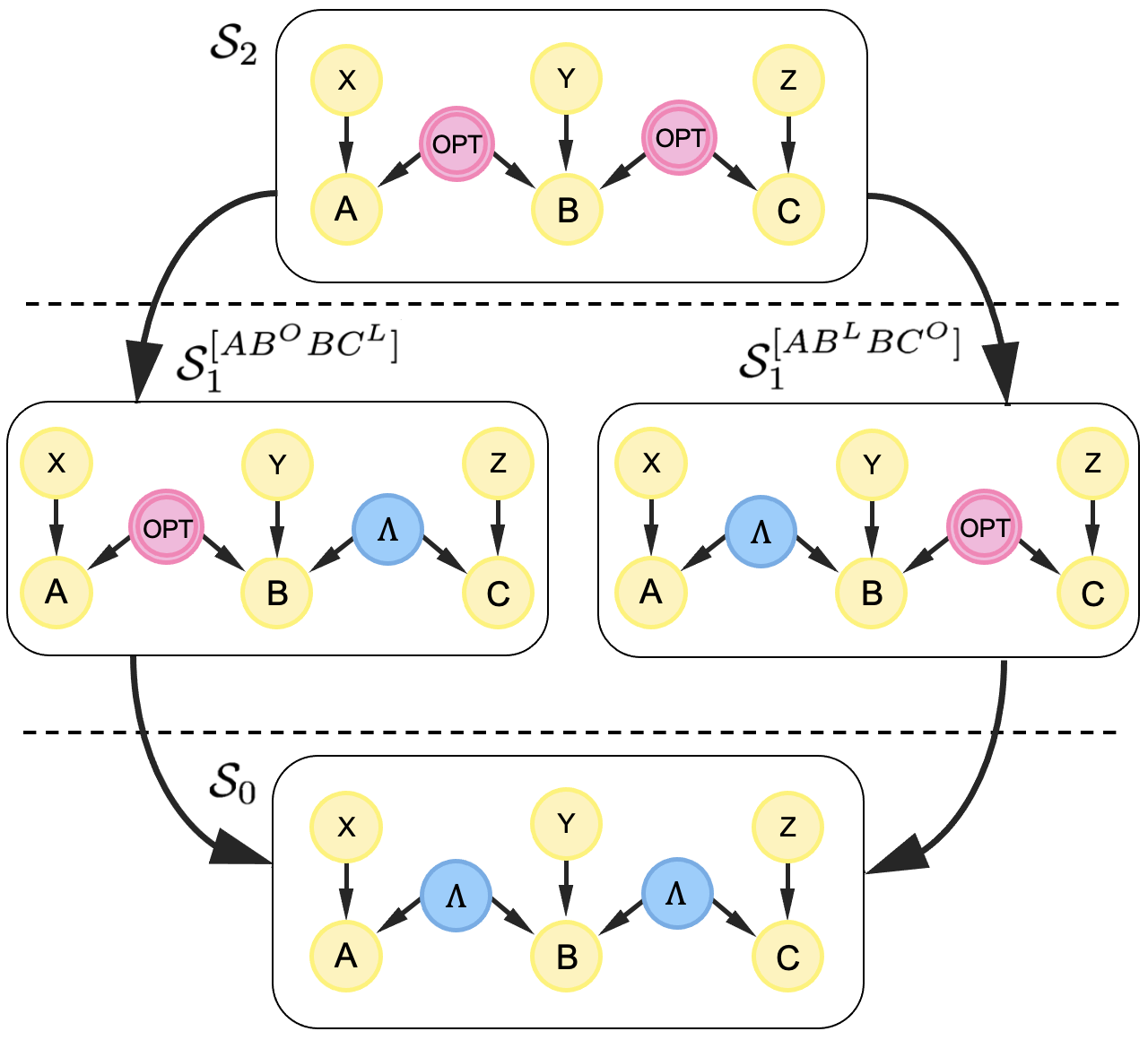}
\caption{Representation of the observational causal order for the 3-chain network. Outside arrows represent dominance (i.e. the set of correlations compatible with one scenario are included in the set of correlations compatible with the other). $\Lambda$ represents a classical source, while $OPT$, an OPT one.}
\label{fig:causal_order}
\end{figure}

Let us consider, for simplicity, the 3-chain scenario with three parties and two bipartite shared sources (Fig.~\ref{fig:ex_bilocality}). We begin by recognizing that the 3-chain scenario induces different sets of correlations based on the nature of the sources therein (Fig.~\ref{fig:causal_order}). 
Furthermore, these sets of correlations are partially-ordered by set inclusion.
First, we define
the set $\mathcal{S}_0$ as the correlations compatible with the 3-chain scenario using only classical sources, that is, those that can be expressed as follows:
\begin{align}\label{eq:def_PS1}
\begin{split}
     p_{\mathcal{S}_0}(a,b,c|&x,y,z) 
     \\ 
      = \displaystyle\int &d\lambda_1 d\lambda_2 p(\lambda_1)p(\lambda_2)\times...
     \\ &p(a|x,\lambda_1) p(b|y,\lambda_1, \lambda_2) p(c|z,\lambda_2) \,.
\end{split}\end{align}
Additionally, one can define $\mathcal{S}_1^{[AB^OBC^L]}$ and $\mathcal{S}_1^{[AB^LBC^O]}$, representing the sets of correlations compatible with one source being classical and the other being OPT-compatible. An OPT-compatible source refers to the most general  \emph{Operational Probabilistic Theory}~\cite{chiribella2010probabilistic} boxes that satisfy no-signalling constraints, device-replication and source independence~\cite{gisin2020constraints}. The sets $\mathcal{S}_1^{[AB^OBC^L]}$ and $\mathcal{S}_1^{[AB^LBC^O]}$ are illustrated in Fig. \ref{fig:causal_order} and can be expressed
in the following way:
\begin{subequations}\label{eqs:halflocal}
\begin{align}
\begin{split}
     &p_{\mathcal{S}_1^{[AB^OBC^L]}}(a,b,c|x,y,z)\\
    &=\int d\lambda\; p(\lambda) p_{{}^{AB^O}}(a,b|x,\{y,\lambda\})p(c|z,\lambda)\;,
\end{split}\shortintertext{and}
\begin{split}
    &p_{\mathcal{S}_1^{[AB^LBC^O]}}(a,b,c|x,y,z)\\
    &=\int d\lambda\; p(\lambda)p(a|x,\lambda)p_{{}^{BC^O}}(b,c|\{y,\lambda\},z)\;,
\end{split}
\end{align}
\end{subequations}
where in Eqs.~\eqref{eqs:halflocal} $p_{AB^O}$ and $p_{BC^O}$ each denote a respective bipartite set of no-signalling correlations and where $\lambda$ always serves as an input for $B$. Notice that, in the bipartite case, the NS corresponds to the most general OPT boxes.

\begin{figure}[b]
\centering
\includegraphics[width=6.5cm]{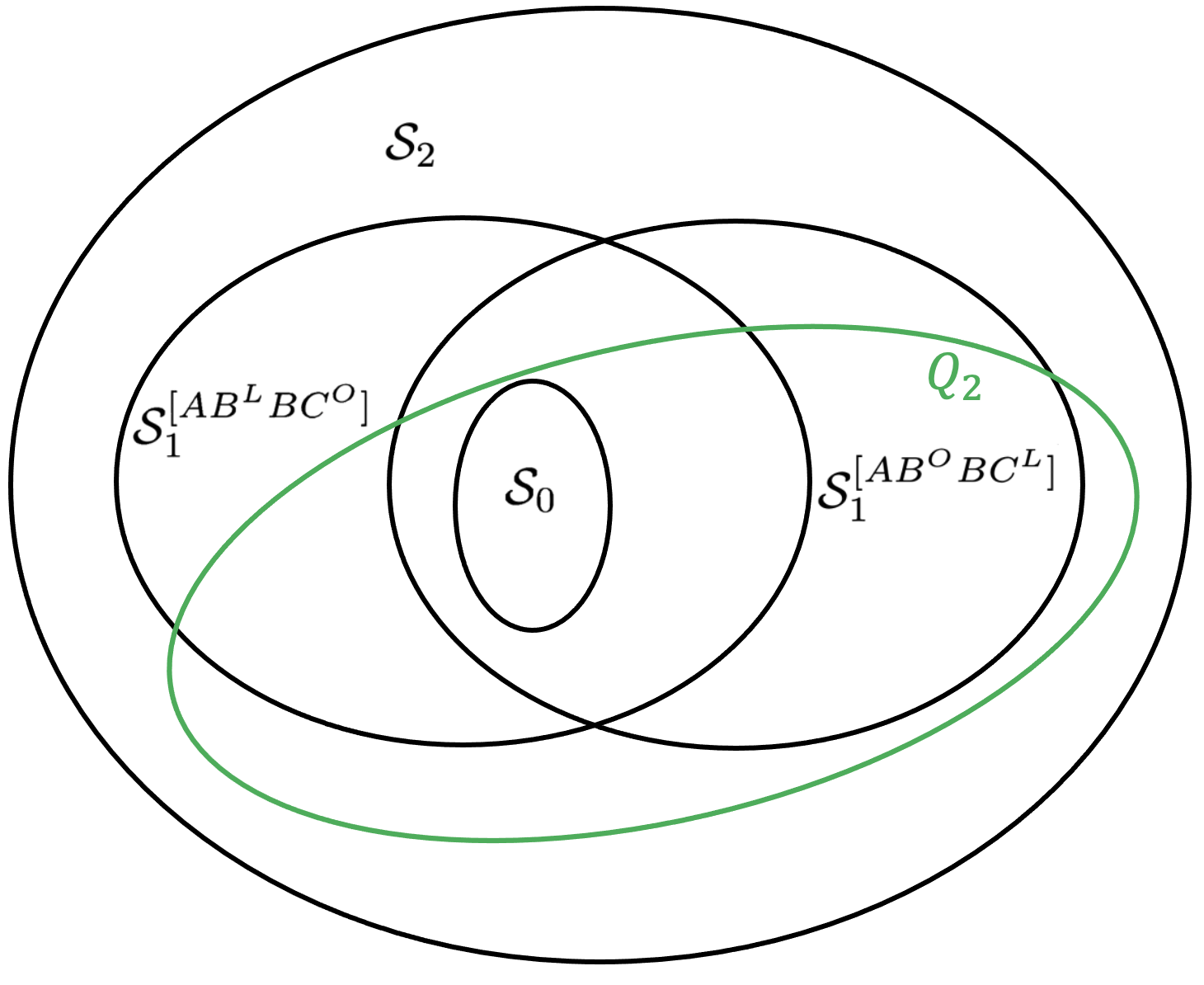}
\caption{Venn diagram representation of the sets of correlations realizable by the corresponding networks depicted in Fig.~\ref{fig:causal_order}. Note that ${\mathcal{S}_2 \displaystyle\supset \left(\mathcal{S}_1^{[AB^OBC^L]} \cap \mathcal{S}_1^{[AB^LBC^O]}\right)}$ and that ${\mathcal{S}_0 \displaystyle\subset \left(\mathcal{S}_1^{[AB^OBC^L]} \cup \mathcal{S}_1^{[AB^LBC^O]}\right)}$. \emph{Only} the bilocal set $S_0$ is wholly contained within the quantumly-realizable set $Q_2$, i.e. ${S_0 \subset Q_2}$. Note that \emph{none} of the sets in the diagram are actually convex. }
\label{fig:sets}
\end{figure}

Clearly it holds both that ${\mathcal{S}_0\subset \mathcal{S}_1^{[AB^OBC^L]}}$ and that  ${\mathcal{S}_0\subset \mathcal{S}_1^{[AB^LBC^O]}}$, while in general the latter sets are not comparable (cf. Fig.~\ref{fig:causal_order} and \ref{fig:sets}).
Finally, a superset $\mathcal{S}_2$ can be defined as the correlations arising by any OPT in the 3-chain network, 
(cf. Fig.~\ref{fig:causal_order}) 
that is:
\begin{align}
    \nonumber p_{\mathcal{S}_2}&(a,b,c|x,y,z) \text{ s.t. } \\ 
     \sum_b p_{\mathcal{S}_2}&(a,b,c|x,y,z)=p(a|x)p(c|z)\\
    \nonumber \sum_a p_{\mathcal{S}_2}&(a,b,c|x,y,z)=\sum_a p_{\mathcal{S}_2}(a,b,c|x,y,z')\;\forall z,z'\\
    \nonumber \sum_c p_{\mathcal{S}_2}&(a,b,c|x,y,z)=\sum_c p_{\mathcal{S}_2}(a,b,c|x',y,z)\;\forall x,x'
\end{align}

Note that the subindices of $\mathcal{S}_0$, $\mathcal{S}_1$ and $\mathcal{S}_2$ indicate the number of nonclassical sources required in the network to compute the correlations belonging to those sets.


The introduction of these sets allows us to formally define \emph{Minimal Network Nonclassicality} (\mnn),  a form of nonclassicality that is certainly not \CABW, in the 3-chain scenario.
In particular, a correlation $p(a,b,c|x,z)$ is \mnn\ in such network iff it belongs to the set $\{\mathcal{S}_1^{\cap} \setminus \mathcal{S}_0\}$ (cf. Fig.~\ref{fig:sets_results}), where $\mathcal{S}_1^{\cap}\equiv (\mathcal{S}_1^{[AB^OBC^L]} \cap \mathcal{S}_1^{[AB^LBC^O]})$. That is, \mnn correlations are nonclassical but can still be explained by constraining any of the two sources to be classical.



The generalisation for an arbitrary network with $m$ parties and $n$ sources is straightforward. We can similarly define sets of correlations based on the types of sources allowed. Specifically, $\mathcal{S}_0$ and $\mathcal{S}_n$ corresponds to correlations compatible with all sources being classical or all sources being nonclassical, respectively. Additionally, we can define $\mathcal{S}_i^{\cap}$ and $\mathcal{S}_i^{\cup}$ as, respectively, the intersection and the union of the possible sets of correlations compatible with $i$ nonclassical sources. 

Then, we can formally define \mnn for any network.

\begin{definition}[\textsf{M}inimal \textsf{N}etwork \textsf{N}onlocality (\MNN)]
\label{def:MNN}
In a given network with $m$ parties and $n$ nonclassical sources,  
a correlation $p(\Bar{a}|\Bar{x})$ is considered to be \textsf{M}inimally \textsf{N}etwork \textsf{N}onlocal  (hereafter \mnn) iff it cannot be modeled by allowing all the sources in the network to be classical, while it is compatible with \emph{all} causal interpretations wherein exactly one of the $n$ sources is of OPT-variable nature and the others are classical\footnote{Note that this definition includes the case of hybrid networks where there are both classical and nonclassical sources present in the network. See Appendix~\ref{app: hybrid_networks}.1}. Here, $\Bar{x} = (x_1,...,x_m)$ represents the inputs and $\Bar{a} = (a_1,...,a_m)$, the outputs of each party. In other words, $p(\Bar{a}|\Bar{x})$ is \mnn\ iff it belongs to the set $\mathcal{S}_1^{\cap}\setminus \mathcal{S}_0$.
\end{definition}

Note that the definition of \mnn has been explicitly constructed to ensure the following:
\begin{prop}\label{prop:MNNimpliesnotUNCLE}
If a correlation is minimally network nonclassical with respect to a network with only bipartite \emph{nonclassical} sources\footnote{Note that there can be N-partite \emph{classical} sources.}, then it must \emph{not} rely on any particular subset of parties exploiting a nonclassical common cause.
\\\indent\begin{minipage}[t]{0.93\linewidth}
\textbf{Shorthand:} ${\mnn \implies \lnot \UNCLE}$
\\\indent\textbf{Mnemonic:} Seeing the \textsf{M}\!(\!oo\!)\textsf{\!N\!N} requires an absence of \UNCLE.
\end{minipage}\end{prop}

The proof of this proposition and an example of a correlation that is simultaneously \mnn and \SUN in a network with tripartite nonclassical sources can be found in Appendix~\ref{app: counterex}.

Putting Props.~\ref{prop:CABWimpliesUNCLE} and \ref{prop:MNNimpliesnotUNCLE} together immediately leads us to the following theorem, which essentially powers all of the results that follow:
\begin{theorem}\label{theo:main}
If a correlation is minimally network nonclassical with respect to a network with only bipartite nonclassical sources, then evidently it has escaped the shadow of Bell's theorem.
\\\indent\begin{minipage}[t]{0.93\linewidth}
\textbf{Shorthand:} ${\MNN \implies \lnot \CABW}$
\\\indent\textbf{Mnemonic:} In \textsf{M}\!(\!oo\!)\textsf{\!N\!N}light there are no {\Shadow}s.
\end{minipage}\end{theorem}

Notice that, we are interested in theory-independent notions of network nonclassicality. In the particular case of the 3-chain scenario, for instance, we considered the sets $\mathcal{S}_0$, $\mathcal{S}_1^{[AB^OBC^L]}$, $\mathcal{S}_1^{[AB^LBC^O]}$ and $\mathcal{S}_2$ and the corresponding hierarchy of correlations based on general no-signalling OPTs. Clearly, a corresponding hierarchy can be defined using quantum sources of correlations defining the sets $\tilde{\mathcal{S}}_0 = \mathcal{S}_0$, $\tilde{\mathcal{S}}_1^{[AB^QBC^L]}$, $\tilde{\mathcal{S}}_1^{[AB^LBC^Q]}$ and $\tilde{\mathcal{S}}_2$, for which it holds in general that $\tilde{\mathcal{S}}_i\subset\mathcal{S}_i$.

A few comments are in order: 
The \mnn property describes
correlations whose nonclassicality cannot be traced back to a specific source (any of the sets compatible with one nonclassical source can explain them), and might therefore be interpreted as a ``delocalized'' nonclassicality. 
In this sense, \mnn explicitly avoids \UNCLE strategies when we study networks with only bipartite nonclassical sources. In particular, Theorem~\ref{theo:main} implies that all the Bell-embedding examples discussed above (cf. Appendix~\ref{app:shadow_examples}) are not \mnn. 
Secondly, \mnn should be contrasted with full network nonclassicality (\fnn)~\cite{Pozas_Kerstjens_2022}. \fnn essentially picks out those correlations which are \emph{maximally expensive} in the sense that they can only be realized by taking \emph{all} the latent sources in the network to be nonclassical. In stark contrast, \mnn picks out correlations which are \emph{minimally expensive} in the same sense. That is, the cost of realizing an \mnn correlation can be paid for in so many different ways, such that no \emph{particular} subset of the parties need to share a nonclassical common cause to obtain the correlation. As such, while we here consider \mnn correlations to be unambigously outside of the shadow of Bell's theorem for networks with only bipartite nonclassical sources, the very same correlations would necessarily be assessed as \emph{lacking novelty} in the \fnn framework! While it is tautologically impossible for any correlation to be both \mnn and \fnn, we would reiterate that within the novelty assessment framework advocated here it is perfectly plausible for a correlation to be both \fnn and at the same time untraceable back to Bell's theorem. That is, Thm.~\ref{theo:main} only works in one direction. 


In the case of the 3-chain scenario analyzed here (see Fig. \ref{fig:causal_order}), in which \mnn  corresponds to $\mathcal{S}_1^{\cap} \setminus \mathcal{S}_0$, \fnn would be defined by $\mathcal{S}_2\setminus \mathcal{S}_1^{\cup}$.
Both \fnn\ and \mnn are a properties relative to the fixed network topology, and are not, in general, preserved when modifying the latter. Moreover, the sets of \mnn and \fnn correlations are both generally nonconvex.

In the following we provide instances of \mnn\ correlations realizable in the 3-chain scenario with quantum sources, as well as several other examples showing the nontrivial hierarchy of correlations arising already in this simple network.

\section{\mnn in the simplest network}
\label{sec:results}



In order to analyze the nonclassicality-hierarchy above introduced and to exhibit examples of \mnn correlations, we continue to consider what is arguably the simplest nontrivial network featuring more than one shared source, that is the 3-chain scenario (Fig.~\ref{fig:causal_order}) described in Sec.~\ref{sec:beyond_bell}, in its simplest configuration. That is, we take all outputs $\{a,b,c\}$ to be binary, as well as the inputs $x$ of Alice and $z$ of Charlie, while $y$ is trivial (Bob has no input). In fact, any reduction of these chosen cardinalities trivializes the scenarios (cf. Appendix~\ref{sec:minimal_scenario}). We developed tools to assess the nonclassicality properties of a given correlation $p(a,b,c|x,z)$ in this configuration.
First, to determine whether a correlation is  classical in the minimal 3-chain scenario (i.e. belonging to $\mathcal{S}_0$ as per Eq.~\eqref{eq:def_PS1}) we developed an oracle using the \emph{Gurobi} solver~\cite{gurobi}. The oracle uses a simple algorithm based on a pictorial representation of the simplest 3-chain scenario (see details in Appendix~\ref{bilocal_oracle}). This oracle can also be generalized to larger cardinalities of inputs and outputs, by exploiting the formalism proposed in~\cite{branciard2012bilocal}. 
Secondly, to determine the compatibility with $\mathcal{S}_1^{[AB^OBC^L]}$ and $\mathcal{S}_1^{[AB^LBC^O]}$ (Fig. \ref{fig:causal_order}), Linear Programs (LP) are used. These programs are based on imposing the conditions of no-signalling, source-independence and compatibility with the probability distribution (further details can be found in Appendix \ref{feasibility_LPs}). The results obtained with these LPs are equivalent to the ones obtained using the inflation technique~\cite{Wolfe_2019,boghiu2023inflation}.

\begin{figure}[t]
\centering
\includegraphics[width=6.5cm]{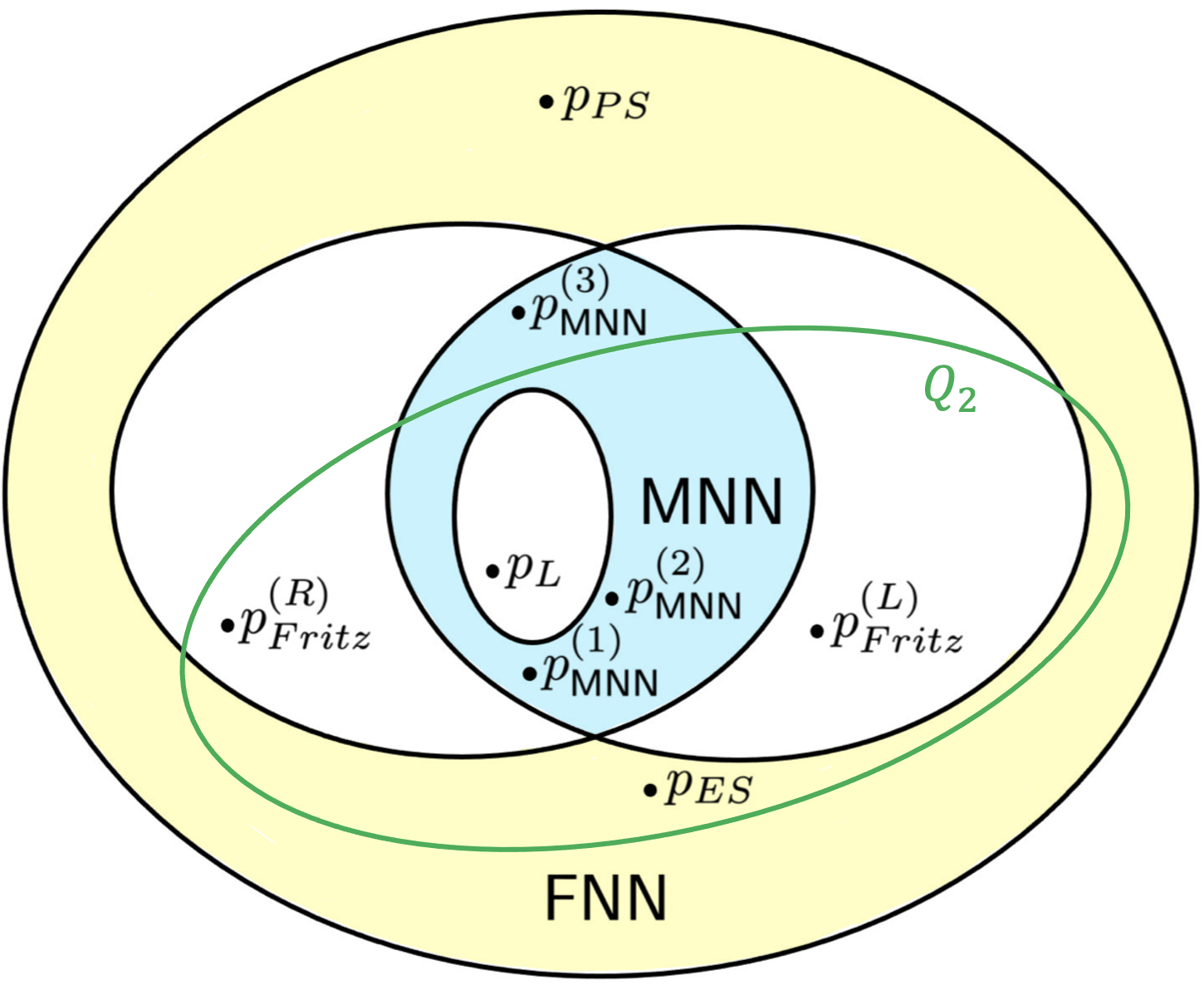}
\caption{We recapitulate the Venn diagram per Fig.~\ref{fig:sets}, where here we shade in the subsets of correlations which are \MNN (blue region) and \fnn (yellow region). The named points within the figure correspond to example correlations in the main text. }
\label{fig:sets_results}
\end{figure}

We are then able to demonstrate the existence of correlations in all the different parts of the Venn diagram, i.e. proving that the causal hierarchy in Fig.~\ref{fig:causal_order} is nontrivial. In Fig.~\ref{fig:sets} we visually depict the different sets of correlations, $S_0$, $S_1^{[AB^OBC^L]}$, $S_1^{[AB^LBC^O]}$, $S_2$ and the set of quantumly-realizable correlations in the 3-chain scenario $\mathcal{Q}_2$. In Fig.~\ref{fig:sets_results} we also depict the \mnn and \fnn sets and the correlations that will be discussed in this section. Using our numerical methods, we were able to show that the \mnn set is non-trivial, and furthermore, realisable within quantum theory.

An example of this novel type of nonclassicality, denoted as $p_{\mnn}^{(1)}$ in Fig. \ref{fig:sets_results}, is found in the simplest 3-chain scenario, thus proving the existence of the gap between $\mathcal{S}_0$ and $\mathcal{S}_1^\cap$. This correlation $p_{\mnn}^{(1)}=\mu p_{ES}+(1-\mu)p_L$ is a mixture of a version of the entanglement-swapping scenario $p_{ES}$ ~\cite{branciard2010characterizing,branciard2012bilocal} and a local Bell test $p_L$ (more details are provided in Appendix~\ref{examples_MNN}). By construction, the entanglement-swapping distribution is feasible using entangled sources and entangled measurements. Indeed, it is a fully-network nonclassical correlation~\cite{Pozas_Kerstjens_2022}. However, when convexly combined with a specific local Bell test, it is not necessary for both sources to be nonclassical. We selected scenarios that ensure the convex combination remains quantum, thereby demonstrating that this type of nonclassicality can manifest within quantum theory. Furthermore, it is noteworthy that the protocol proposed in~\cite{lauand2024quantum} can be modified for use in the 3-chain scenario (see details in Appendix~\ref{examples_MNN}). This leads to a probability distribution, $p_{\mnn}^{(2)}$, which is compatible with quantum theory, depends on a parameter $\theta$ and it is \mnn for $\theta\in(0,\pi/4)\cup(\pi/4,\pi/2) $. Furthermore, we studied the amount of noise that the aforementioned \mnn examples, that are quantumly realizable, could tolerate before becoming classically compatible. This coincides with the point at which incompatibility with the \mnn arises. If each source emits a Werner state~\cite{werner1989quantum} of the form $\rho_v = v \rho + (1-v)\mathbb{1}$, where $v\in[0,1]$ is the visibility and $\rho$ is the state produced by the source when there is no noise, then a quantum demonstration of \mnn can be achieved with appropriate measurements when $v\geq v_{crit}$. The value of the critical visibility for the convex mixture $p_{\mnn}^{(1)}$ is $v_{crit}\approx 0.987$ for the convex weight $\mu=0.65$ (see Appendix \ref{app:shadow_examples}). However, in the case of $p_{\mnn}^{(2)}$ with $\theta=\pi/8$, the critical visibility reach a notable lower value of $v_{crit}\approx 0.861$, thus providing a candidate to experimentally observe \mnn.

Finally, we find post-quantum examples of minimal nonclassical correlations, denoted as $p_{\mnn}^{(3)}$ in Fig. \ref{fig:sets_results}. These are generated using a similar mixing strategy with a post-quantum box $p_{PS}$ and the same local Bell test $p_L$ (further details in Appendix \ref{examples_MNN}). Note that we can certify that these examples are post-quantum via quantum inflation ~\cite{wolfe2021quantum, boghiu2023inflation}.

Furthermore, in order to prove the nontriviality of every set represented in Fig. \ref{fig:sets_results}, we give an example of a correlation $p_{Fritz}^{(R)}$ that is contained in $\mathcal{S}_1^{[AB^OBC^L]}\setminus \mathcal{S}_1^\cap$. That correlation is a wiring-based embedding of the CHSH inequality considered by Fritz~\cite{Fritz_2012} (cf. Sec~\ref{app:shadow_examples}). Similarly, we can construct a correlation $p_{Fritz}^{(L)}$ that belongs to $\mathcal{S}_1^{[AB^LBC^O]}\setminus \mathcal{S}_1^\cap$.

All the results mentioned in this section pertain to the simplest configuration in the 3-chain network and represent the first examples requiring quantum sources in that configuration. Nevertheless, it is noteworthy that the protocol proposed in ~\cite{tavakoli2021bilocal} for the 3-chain scenario with output cardinalities $\{|A|,|B|,|C|\}=\{2,4,2\}$ and input cardinalities $\{|X|,|Y|,|Z|\}=\{3,1,3\}$, which utilizes the elegant joint measurement~\cite{gisin2017elegant}, is also \mnn. This result was computationally demonstrated by employing analogous programs to those described in Appendices \ref{bilocal_oracle} and \ref{feasibility_LPs} but tailored for higher cardinalities.

\section{Motivating \mnn and alternatives}
\label{sec:discussion}

The desideratum motivating this work is a conservative framework for assessing the novelty of nonclassical phenomena arising in nontrivial network topologies, i.e. those with more than one independent nonclassical source.

Recall that a correlation is said to be \UNCLE iff it can be explained without relying on any particular pair of parties exploiting a nonclassical common cause (notice that here we do not distinguish between a nonclassical common cause which is originally present in the network as a source from one which is induced via postselection). Recall further that ${\MNN \implies \lnot \UNCLE \implies \lnot \CABW}$\maria{\footnote{Notice that in this discussion, we restrict ourselves to the case of networks with only bipartite nonclassical sources.}}. However, perhaps a correlation need \emph{not} be \MNN in order to be escape Bell theorem's \textsf{Shadow}? We will answer this in the affirmative in Prop.~\ref{prop:MNNrelaxation}. Anticipating that finding, we herein consider what sort of \emph{relaxation} of \MNN would nevertheless still imply  $\lnot \UNCLE$.

To be clear, however, we leave as an open question whether or not ${\UNCLE \implies \CABW}$. We suspect (but cannot be certain) that as a community we will eventually define \CABW in a manner that excludes certain \fnn correlations in addition to excluding all \mnn correlations.


A superficially appealing --- but ultimately mistaken --- approach for constructing a superset of \mnn correlations while retaining the ${\lnot\UNCLE}$ implication would be to include all nonclassical correlations which can be explained every which way of constraining a single source within the network to be classical. Such a correlation evidently does not rely on exploiting the nonclassicality of a single source \emph{originally present} in the network. 
The hypothetical set of such correlations formally corresponds to $\mathcal{S}_{n-1}^{\cap}\setminus \mathcal{S}_{0}$(cf. Sec.~\ref{sec:beyond_bell}).
The astute reader may anticipate the fallacy in the above construction: despite requiring multiple qualitatively distinct causal explanation for the same correlation, the aforementioned criterion leaves open the possibility of embedding the violation of a Bell inequality via an \emph{induced} common cause. 

To illustrate the problem with a concrete example, consider the correlation in the triangle scenario described in Sec.~\ref{sec:beyond_bell}. As explained there, this correlation is based on the entanglement-swapping protocol, resulting in a correlation that is not merely \UNCLE, but is actually pretty clearly \CABW as the correlation violates the CHSH inequality between Alice and Charlie under suitable postselection of Bob's outcome. At the same time, this correlation turns out to be compatible with $\mathcal{S}_2^{\cap}$ (cf. App.~\ref{app: triangle_causal_hierarchy}), thereby highlighting the insufficiency of imposing \emph{merely} compatibility with $\mathcal{S}_2^{\cap}$ instead of truly imposing $\lnot \UNCLE$.

We thus tailored the definition of \MNN to ensure that it would not admit any \UNCLE correlation. At the same time, we acknowledge that alternatives may also be considered. For instance, we could prohibit the use of entangled measurement in conjunction with the aforementioned compatibility with every one-classical source configuration. It should be noted that such a conjoined-restrictions criterion would exclude the entanglement-swapping example previously mentioned. Nonetheless, we opted for \MNN as it is more readily formalizable and verifiable. 

\begin{figure}[t]
\centering
\includegraphics[width=7cm]{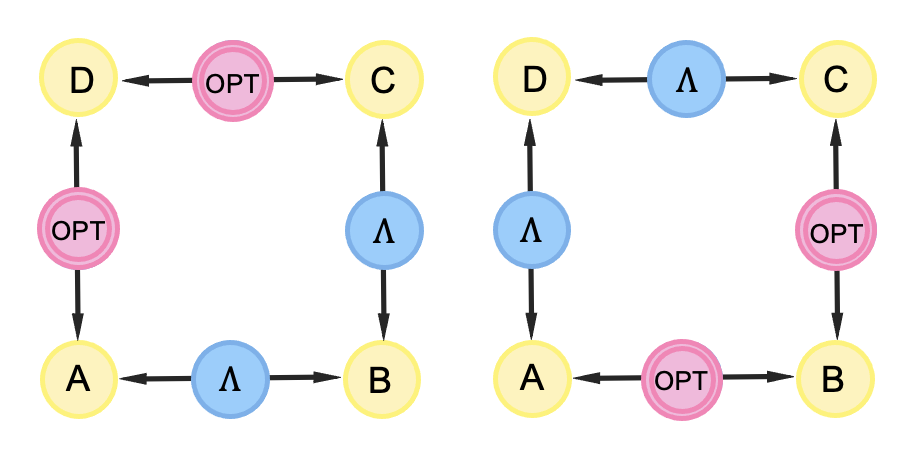}
\caption{Representation of two possible configurations of the square network with two classical sources and two nonclassical ones. The intersection of correlations feasible in both configurations defines a set $\bar{\mathcal{S}}$.}
\label{fig:square}
\end{figure}

We thus acknowledge the potential restrictiveness of the \mnn notion. Consider, for instance, the square network depicted in Fig.~\ref{fig:square}. In such scenario one can define the set of correlations, say $\bar{\mathcal{S}}$, that are compatible with both configurations represented (that is, classical sources shared by AB and BC and nonclassical sources shared by CD and DA (left), or the complementary configuration (right)). Clearly, \mnn correlations are in principle a strict subset of $\bar{\mathcal{S}}$.
At the same time, it is not difficult to see that correlations belonging to $\bar{\mathcal{S}}$ prevented to feature \UNCLE nonclassicality (and, therefore, \CABW nonclassicality) leveraging nonclassicality of any of the four network source, as well as of sources induced via entanglement swapping. Another relaxation of the \mnn definition that would also imply $\neg$\UNCLE would be to require compatibility with more than one explanation in which only one of the sources is nonclassical, rather than requiring compatibility with all such explanations, as is done in \mnn.

In summary, we can state the following proposition:

\begin{prop}\label{prop:MNNrelaxation}
The fact that a correlation is not minimally network nonclassical does not imply that it relies on any particular subset exploiting a nonclassical common cause. Equivalently, not relying on any particular subset exploiting a nonclassical common cause to explain a correlation does not mean that such correlation is minimally network nonclassical.
\\\indent\begin{minipage}[t]{0.93\linewidth}
\textbf{Shorthand:} ${\lnot \MNN \centernot\implies \UNCLE}$.
\\\textbf{Mnemonic:} \UNCLE can be absent despite no \textsf{M}\!(\!oo\!)\textsf{\!N\!N}.
\end{minipage}
\end{prop}

\noindent\begin{minipage}{\textwidth}
\noindent\subsection{Epigraph}
{\itshape
To recap our new proving scheme,
\\recall {\Shadow}s need \SUN to be seen,
\\\indent\hspace{1em}\begin{minipage}[t]{\textwidth}
so if {\Shadow}s you dread,
\\seek the \textup{\textsf{M}\!(\!oo\!)\textsf{\!N\!N}} overhead
\end{minipage}\vskip0.2em
as no \SUN is in sight when the \textup{\textsf{M}\!(\!oo\!)\textsf{\!N\!N}} beams.
}
\end{minipage}
\newpage
\section{Conclusions}
In this work, motivated by the goal of finding novel types of statistical nonclassicality in (quantum and postquantum) networks,
we introduced the notion of Minimal Network Nonclassicality (\mnn), describing correlations whose nonclassicality can be explained via the use of a single nonclassical source in any point of the given network. This property guarantees, at the same time, that such nonclassicality is non-expensive in terms of nonclassical sources, and cannot be traced back to obvious liftings of simpler Bell nonclasscal scenarios (when talking about networks with only binary nonclassical sources).
Via numerical results we prove the existence of \mnn, both in the quantum and in the post-quantum regime, already in the simple 3-chain (or bilocality) scenario with minimum cardinality of inputs and outputs. In particular, we find a point ($p_\mnn^{(2)}$, cf. Sec.~\ref{sec:results}) with specific noise-resistance of $\sim 14\%$, thus providing a candidate of experimentally-observable \mnn. It is important to stress that we introduced the notion of \mnn\ as a \emph{sufficient} condition for a nonclassical correlation to be genuinely beyond simple \CABW nonclassicality, however we do not claim such condition to be necessary, nor to be \emph{the} definition of genuine network nonclassicality (cf. Discussion~\ref{sec:discussion}). Rather, it highlights a peculiar phenomenon which might be useful in understanding novel properties and applications of quantum causal networks.

Finally, while major efforts have been recently focused on characterizing the simplest possible examples of nonclassicality in the triangle network~\cite{renou2019genuine, boreiri2023towards}, we notice that
the 3-chain scenario, in its minimal-cardinality configuration,
arguably represents the simplest network scenario with more than a single-source (and without direct causal influence among parties, cf. App.~\ref{sec:minimal_scenario}). In this work we served a thorough analysis of this simple network
and developed the tools necessary to solve the feasibility problem in the different sets of the nonclassicality hierarchy. See Fig.~\ref{fig:sets_results}  for a summary of pertinent results. 
%
%
%
%
Further future directions and challenges include the exploration of the \mnn\ set in larger networks, 
as well as proof-of-principle implementation.

\onecolumn
\clearpage
\section{Acknowledgements}

We thank Pedro Lauand, Daniel Centeno, Marina Maciel Ansanelli, Victor Gitton and Marc-Olivier Renou for discussions and comments. 
\begin{wrapfigure}[6]{R}{2cm}
\includegraphics[width=\linewidth]{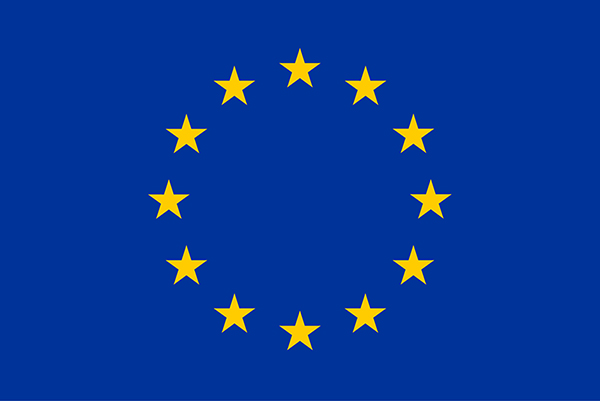}
\end{wrapfigure}
This work is supported by the Spanish Ministry of Science and Innovation MCIN/AEI/10.13039/501100011033 (CEX2019-000910-S, PRE2019-088482),  Fundació Cellex, Fundació Mir-Puig, and Generalitat de Catalunya through the CERCA program. 

Research at Perimeter Institute is supported in part by the Government of Canada through the Department of Innovation, Science and Economic Development and by the Province of Ontario through the Ministry of Colleges and Universities. 
P.A. is supported by the QuantERA II programme, that has received funding from the European Union’s Horizon 2020 research and innovation programme under Grant Agreement No 101017733, and from the Austrian Science Fund (FWF), project I-6004.

\section*{Authors contributions}
M.C.A. performed the analytical and numerical analysis and wrote the first draft of the manuscript. E.-C.B. contributed to the computational implementations. P.A. and E.W. conceived the project and supervised the research. All authors contributed to the discussions and to the final version of the manuscript.
\setlength{\bibsep}{.4\baselineskip plus .1\baselineskip minus .1\baselineskip}

\bibliography{BIB.bib}


\appendix
\clearpage
\section{\CABW and related inequalities}
\label{app:shadow_examples}

Here we detail simple known embeddings of bipartite Bell nonclassicality and the relative inequalities that arise from those scenarios.\\
\subsection{Post-selection-based nonclassicality}
\label{subsection:E-S}
Entanglement-swapping is a well-known phenomenon that generates nonclassicality in the 3-chain scenario~\cite{branciard2010characterizing, branciard2012bilocal}. It involves establishing nonclassical correlations between two particles that have never interacted previously. Let us consider the scenario depicted in Fig.~\ref{fig:ex_bilocality}. The sources emit pairs of particles in a maximally entangled state, say $\ket{\phi^+} = (\ket{00}+\ket{11})/\sqrt{2}$. Bob performs a coarse-grained Bell state measurement on the two received particles, yielding two possible outputs $b=0,1$, which correspond to the measurement operators $\hat{B}_0 = \ket{\psi^+}\bra{\psi^+}$ and $\hat{B}_1 = \mathbb{1} - \ket{\psi^+}\bra{\psi^+}$, respectively. Then, when Bob outputs 0, he performed entanglement swapping, and Alice and Charlie will be sharing a maximally entangled state. On the other hand, Alice and Charlie perform the measurements in a way that when Bob outputs 0, they can violate the CHSH inequality. Specifically, they can perform measurements that maximally violate the CHSH inequality, which are $\hat{A_0}=(\sigma_x-\sigma_z)/\sqrt{2}$, $\hat{A_1}=(\sigma_x+\sigma_z)/\sqrt{2}$, $\hat{C_0} = \sigma_x$ and $\hat{C_1} = \sigma_z$. In particular, the partial probability distribution over those events where $b=0$ is given by:
\begin{align}\label{eq:CHSH_after_postselection}
\begin{array}{c|cccc}
(a,b,c) \backslash (x,z) & (0,0) & (0,1) & (1,0) & (1,1) \\
\hline
(+1,0,+1) & p_+ & p_- & p_+ & p_+ \\
(+1,0,-1) & p_- & p_+ & p_- & p_- \\
(-1,0,+1) & p_- & p_+ & p_- & p_- \\
(-1,0,-1) & p_+ & p_- & p_+ & p_+ \\
\vdots
\end{array}
\qquad\mbox{where} \quad p_{\pm} = \frac{2 \pm \sqrt{2}}{32}\,.
\end{align}

It is sufficient to test CHSH inequality to see whether entanglement-swapping has occurred. Therefore, we can attribute the violation of the 3-chain classicality to the violation of CHSH between Alice and Charlie when post-selecting on $b=0$. 


\subsection{Wiring-based embedding nonclassicality}
Another method for producing nonclassicality, considered in ~\cite{Fritz_2012}, involves Bob and Charlie sharing a classical source $\lambda$ that can take the value 0 or 1. Bob’s measurement is determined by $\lambda$, while Charlie outputs $\lambda$ and ignores $z$. This can be interpreted as an scenario in which Charlie’s output becomes Bob’s input. Furthermore, conditioning on Charlie, Alice and Bob perform measurements that violates the CHSH inequality, i.e. $\hat{A_0}=(\sigma_x-\sigma_z)/\sqrt{2}$, $\hat{A_1}=(\sigma_x+\sigma_z)/\sqrt{2}$, $\hat{B_0} = \sigma_x$ and $\hat{B_1} = \sigma_z$. Thus, by defining $p_z(a,b|x,c)=p(a,b|x,c,z) = \frac{p(a,b,c|x,z)}{p(c|z)}$, the following necessary inequality is obtained:

\begin{align}
    -2\leq \sum_{a,b} (-1)^{a+b}\left( \frac{p(a,b,0|0,z)}{p(c=0|z)} + \frac{p(a,b,1|0,z)}{p(c=1|z)} + \frac{p(a,b,0|1,z)}{p(c=0|z)} - \frac{p(a,b,1|1,z)}{p(c=1|z)} \right) \leq 2,
\end{align}
which simply corresponds to the CHSH applied to $p_z(a,b|x,c)$. In this manner, we can trace back all the nonclassical correlations to violations of the previously mentioned CHSH inequality. Notice that this protocol is compatible with having one nonclassical source between A and B and a classical source between B and C but not the reverse configuration, as a classical source between A and B prohibits $p_z(a,b|x,c)$ from violating the CHSH inequality.

It is important to note that this represents only one possible wiring configuration; however, numerous alternatives, such as choosing $c=\lambda \oplus z$, yield similar inequalities. Additionally, a symmetric protocol can be considered by permuting A and C so that we have a classical source between A and B and $p_x(b,c|a,z)$ violates CHSH. \\

\subsection{Entanglement-swapping in the triangle}

Consider the same probability distribution as described in Eq.~\eqref{eq:CHSH_after_postselection} but now in the triangle scenario (where all three parties have binary outcomes, and the two parties $A$ and $C$ additionally have binary inputs). In the case of the triangle, this probability distribution admits two radically distinct causal explanations. The commonality between the two explanations, however, is what leads us to conclude that -- with respect to the triangle scenario -- this probability distributed is \CABW.

The first explanation consists of ignoring the source between $A$ and $C$ and then performing the same protocol as in the 3-chain scenario (Sec.~\ref{subsection:E-S}). In this explanation, the CHSH violation between Alice and Charlie conditioned on Bob's outcome is achieved via postselection on successful entanglement swapping by Bob. Hence, this explanation relies on the \emph{induced} nonclassical common cause that appears when conditioning on a particular outcome of the middle party. 

The second explanation of the same correlation utilizes the nonclassical common cause between Alice and Charlie while ignoring the other two sources, exactly the opposite of the first explanation. In this second explanation, the protocol for realizing the correlation is to take the source between $A$ and $C$ to be, say, $\ket{\phi^+} = (\ket{00}+\ket{11})/\sqrt{2}$ and to have $A$ and $C$ perform the measurements that violate the CHSH inequality, $\hat{A_0}=(\sigma_x-\sigma_z)/\sqrt{2}$, $\hat{A_1}=(\sigma_x+\sigma_z)/\sqrt{2}$, $\hat{C_0} = \sigma_x$ and $\hat{C_1} = \sigma_z$. Note that, although this is a quantum protocol, this is not generally the case, as we only require the source to be nonclassical. This second explanation, however, relies on the nonclassicality of a nonclassical source that is \emph{originally present} in the network.

\section{Nonclassical paths}
\label{app: nonclassical_paths}
There are two distinct senses in which one may say that $A$ and $B$ rely on a nonclassical common cause to become correlated: either the nonclassicality arises from a common cause that connects $A$ and $B$ (i.e., a common cause that is already present in the causal structure), or from a common cause that is \emph{induced} by post-selection on (i.e., conditioning on the values of) some other observed variable(s). We now introduce the notion of a \emph{nonclassical path} in order to capture both originally-present and postselection-induced nonclassical common causes in a unified framework. 

\begin{figure}[h]
\centering
        \begin{subfigure}[b]{0.3\textwidth}
                \centering
                \includegraphics[width=0.8\linewidth]{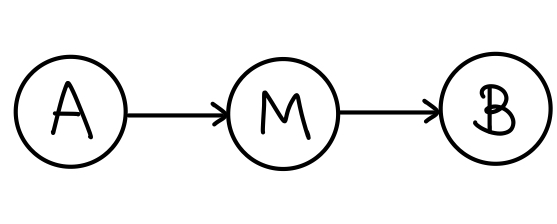}
                \caption{Chain}
                \label{fig:app:chain}
        \end{subfigure}%
        \begin{subfigure}[b]{0.3\textwidth}
                \centering
                \includegraphics[width=0.8\linewidth]{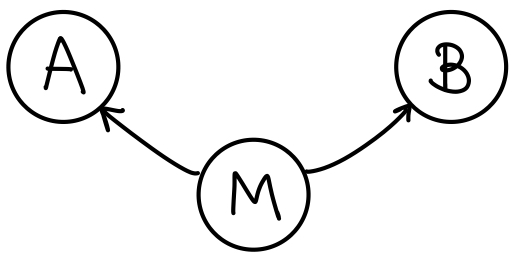}
                \caption{Fork}
                \label{fig:app:fork}
        \end{subfigure}%
        \begin{subfigure}[b]{0.3\textwidth}
                \centering
                \includegraphics[width=0.8\linewidth]{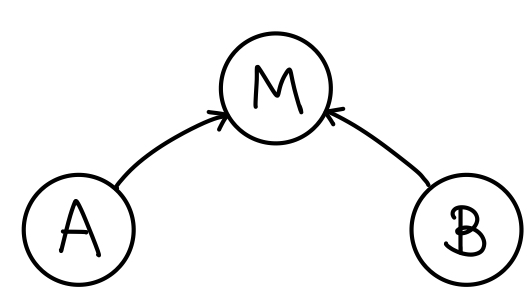}
                \caption{Collider}
                \label{fig:app:collider}
        \end{subfigure}%
        \caption{Representation of the three different causal substructures representing sequential edge pairs along a path.}
\end{figure}

\begin{definition}[Nonclassical path]
    Two variables are said to be path-connected if there is a sequence of edges (direction irrelevant) in the DAG which connects them. There can be multiple paths between the same two variables. A path between two observed variables is said to be \emph{nonclassical} if and only if its sequential edge pairings (namely: chains, forks, and colliders) satisfy the following properties and moreover the path contains at least one fork:
\begin{itemize}
    \item In every chain-type or fork-type sequential edge pair (depicted in Figs.~\ref{fig:app:chain} and \ref{fig:app:fork}) the middle node is a \emph{nonclassical latent node}.\footnote{If the middle node of the chain or fork is classical it constitutes a classical bottleneck, rendering the entire path classical.}
    \item In every collider-type sequential edge pair (depicted in Fig.~\ref{fig:app:collider}) the middle node is a \emph{classical observed node}.
\end{itemize}
\end{definition}

Note that the paths are scenario specific. Notice also that whether or not a path is nonclassical depends on the \emph{particular causal explanation} being invoked to explain the observed distribution. The same path may be nonclassical in some explanations but classical in others. The concept of nonclassical paths is relevant because we claim that a necessary condition for a distribution to be \CABW is if some \emph{fixed} pair of parties is connected by a nonclassical path in \emph{all} causal explanations for the given distribution. 

A path consisting of a single edge between two observed variables is always a classical path. A nonclassical path without any collider-type elements in it should be interpreted as a causal explanation relying on an \emph{originally present, physical} nonclassical common cause. By contrast, a nonclassical path comprising one or more colliders can potentially constitute a \emph{postselection-induced} nonclassical common cause. The notion of a nonclassical path allows us to formalize the concept of \UNCLE.
\begin{definition}[Subset Unavoidable Nonclassicality (\UNCLE), Formal]
A distribution is said to exhibit \UNCLE with respect to a given network if there is some pair of parties such that these parties are connected by at least one \emph{nonclassical} path in every causal explanation capable of reproducing that distribution in that network.\footnote{Here we speak of paths between \emph{parties} in order to use more familiar language, but really, every observed variable should be treated as a party for the analysis of \UNCLE.
}
\end{definition}

Equivalently, a distribution is said to have subset \emph{avoidable} nonclassicality (i.e., $\lnot\UNCLE$) iff for \emph{every} pairing of two parties we can craft a custom causal explanation of the correlation so that the target pair of parties are \emph{not} connected by any nonclassical path (i.e., only classical paths or no path at all). If such a bespoke causal explanation can be tailored for each pair of parties, the correlation is $\lnot\UNCLE$. If there is a pair of parties preventing such a construction --- i.e., a pair of parties that necessitate a nonclassical path connecting them no matter how creative we might get with the causal explanation --- then the correlation is \UNCLE.

\section{Hybrid networks}
\label{app: hybrid_networks}
All definitions in this paper have been formulated to be network-dependent. In particular, the analysis in the main text, together with the proposed examples, has been carried out by considering only the topology of the network. These definitions can be generalized to hybrid networks, in which not only the topology but also the nature of the sources is specified; that is, some sources are fixed to be classical while others are nonclassical. Note that this generalization allows us to study networks with classical shared randomness.

In the case of SUN, the definition extends straightforwardly. For MNN, we say that a correlation is MNN with respect to a given hybrid network if it can be explained using only one nonclassical source, and if that nonclassical source can be placed at any of the locations designated as nonclassical in the hybrid network.

\section{Proof of Proposition ~\ref{prop:MNNimpliesnotUNCLE}}
\label{app: counterex}

For a correlation to exhibit \SUN, there must exist one pair of parties that is connected by a nonclassical path in every possible causal explanation. For a correlation that is \MNN, in every causal explanation involving just a single nonclassical common cause, the only nonclassical paths in that explanation are between those parties that share the sole nonclassical common cause. 
With only bipartite sources, each pair of parties shares at most one potentially nonclassical common cause. As such, the fact that the identity of the nonclassical common cause is fluid (per \MNN) is incompatible with the premise of a single pair of parties being connected by a nonclassical path in every possible causal explanation (per \SUN). This proves that the two properties are mutually exclusive in bipartite networks.

\begin{figure}[ht]
  \centering
  \begin{subfigure}[b]{0.35\textwidth}
    \centering
    \includegraphics[width=\linewidth]{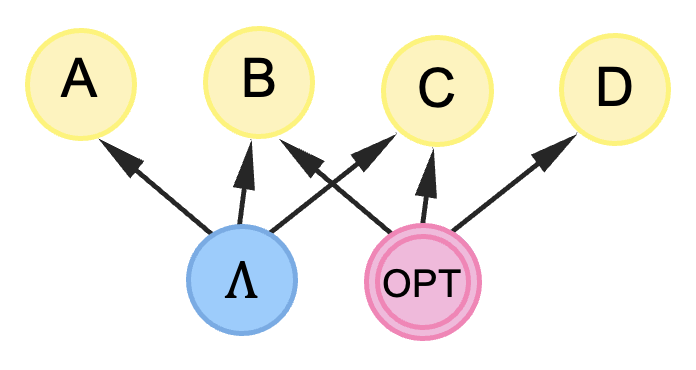}
    \caption{}
    \label{fig:4line_1}
  \end{subfigure}
  \hspace{0.03\textwidth}
  \begin{subfigure}[b]{0.35\textwidth}
    \centering
    \includegraphics[width=\linewidth]{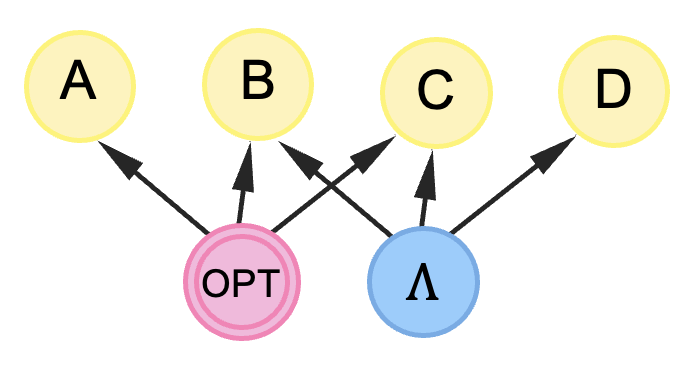}
    \caption{}
    \label{fig:4line_2}
  \end{subfigure}
  \caption{Representation of the two possible configurations of the 4-line scenario with two tripartite sources, where one is classical and the other is nonclassical.}
  \label{fig:4line}
\end{figure}

There are many examples of networks in which one can find correlations that satisfy \mnn and yet exhibit \UNCLE simultaneously when, for instance, tripartite sources are allowed. A simple but illustrative example is provided by the network shown in Fig.~\ref{fig:4line}. 

Consider a probability distribution $p(A,B,C,D)$ such that $p(B,C \mid A,D)$ maximally violates the CHSH inequality. This probability distribution is compatible with a model where one of the sources is classical and the other one is nonclassical, regardless of where the nonclassical source is placed; that is, it is compatible with both of the two models depicted in Fig.~\ref{fig:4line_1} and \ref{fig:4line_2}. Therefore, it is indeed \MNN. Nevertheless, in both explanations the parties $B$ and $C$ exploit the nonclassicality of a single source --- i.e., $B$ and $C$ are connected by a nonclassical path --- so this correlation also exhibits \SUN.

Indeed, in this example it is clear that the correlation is \Shadowed, and hence that the definition of \mnn only implies being outside the shadow of Bell in networks with bipartite sources.

\section{Minimality of the scenario}
\label{sec:minimal_scenario}

In order to study scenarios beyond standard Bell games~\cite{brunner2014bell} (1 shared source, $N\geq 2$ parties), a network should involve at least two independent sources, and (therefore) at least three parties. The smallest nontrivial networks in this sense are the 3-chain-network considered in this work~\cite{branciard2010characterizing,branciard2012bilocal} (sometimes called \emph{bilocal network} in the literature) and the \emph{triangle network} (three parties sharing three bipartite source~\cite{renou2019genuine,tavakoli2022bell})\footnote{Notice that here we are only dealing with networks in which no direct causal influence is allowed between any two parties. For an example of quantum nonclassicality in a 3-party network with 1-way communication from $B$ to $A$ and $C$, see e.g.~\cite{lauand2024quantum}.}. In recent years the latter has attracted a lot of interest~\cite{abiuso2022single,suprano2022experimental,pozas2023proofs}, as it also provides quantum examples of network nonclassicality without random measurement inputs, and research efforts are dedicated towards the characterization of the simplest (i.e., with minimal cardinality of the outputs) games featuring (quantum) nonclassicality in the triangle network~\cite{boreiri2023towards,pozaskerstjens2023postquantum}.
Little attention however has been given to the 3-chain scenario, which is arguably ``smaller". It is in fact, minimal in terms of the number of sources and parties that form the network (2 sources and 3 parties) beyond the standard Bell scenarios.

Motivated by the quest of minimal scenarios exhibiting quantum nonclassicality, we now argue that among all configurations (input-output cardinalities) of the 3-chain network, the case with $2$-$1$-$2$ inputs (indicating the respective cardinalities $|X|$-$|Y|$-$|Z|$) and $2$-$2$-$2$ outputs (respective cardinalities $|A|$-$|B|$-$|C|$) is minimal, making it arguably the truly minimal configuration of beyond-Bell nonclassicality.
Specifically, we consider
\begin{align}
    p(a,b,c|x,z)\quad \text{with dichotomic variables } \{a,b,c,x,z\}\;,
\end{align}
according to the 3-chain network configuration  with trivial input for Bob (Fig.~\ref{fig:ex_bilocality}, Fig.~\ref{fig:causal_order}). In the case of classical theory the decomposition of the corresponding $p$ is then given by
\begin{align}
     p(a,b,c|x,z) = \int_0^1 \int_0^1 d\lambda_1 d\lambda_2 p_A(a|x,\lambda_1) p_B(b|\lambda_1,\lambda_2) p_C(c|z,\lambda_2).
\end{align}

\paragraph{Number of inputs and outputs is minimal.}
If one reduces the cardinality of any of the outputs, effectively the scenario collapses to a 2 party scenario: 1) of the form 
    \begin{align} p(a,c|x,z)=p(a|x)p(c|z)\;, \end{align} 
    which can be fully filled with local strategies; or 2) of the form 
    \begin{align} p(a,b|x,z)=p(a,b|x)\;,
    \end{align}
    with $b$ independent from $x$. It follows that
    \begin{align} p(a,b|x,z)=p(b)p(a|b,x)\;,
    \end{align}
    which, as a set can be filled via classical strategies where the common source between Alice and Bob distributes $b$ according to $p(b)$.

If we reduce the cardinality of any of the two inputs, we obtain a probability of the form \begin{align}
    p(a,b,c|x)     
    \end{align}
    This can be nonclassical iff ${p(a,b|x,c)}$ is nonclassical: if the former is classical according to 
    \begin{align}
    p(a,b,c|x)=\int_0^1 \int_0^1 d\lambda_1 d\lambda_2 p_A(a|x,\lambda_1) p_B(b|\lambda_1,\lambda_2) p_C(c|\lambda_2),
    \end{align}
    it follows immediately that 
    \begin{align}\begin{split}
    p(a,b|x,c)&=\int_0^1 \int_0^1 d\lambda_1 d\lambda_2 p_A(a|x,\lambda_1) p_B(b|\lambda_1,\lambda_2) p(c)^{-1}p_C(c|\lambda_2)
    \\&= \int_0^1 d\lambda_1 p_A(a|x,\lambda_1)p'_B(b|c,\lambda_1)
    \end{split}\end{align}
    is classical in the usual bipartite Bell scenario. Conversely, given a classical 
    \begin{align}
    p(a,b|x,c)=\int_0^1d\lambda_1  p_A(a|x,\lambda_1)p_B(b|c,\lambda_1),
    \end{align}
    one can reproduce 
    \begin{align}
    p(a,b,c|x)\equiv p(a,b|x,c)p(c)=\int_0^1 \int_0^1 d\lambda_1 d\lambda_2 g(\lambda_2) p_A(a|x,\lambda_1) p_B(b|\lambda_1,\lambda_2) p_C(c|\lambda_2)
    \end{align}
    by sharing a classical variable $\lambda_2$ between B and C having its distribution $g$ identical to $p(c)$, and local response function ${p_C(c|\lambda_2)=\delta_{c,\lambda_2}}$.

\section{Oracle for assesing nonclassicality in the 3-chain scenario}
\label{bilocal_oracle}
To construct an oracle for assesing nonclassicality in the 3-chain scenario, we propose a simple formalism based in a pictorial representation.  This approach applies specifically to the simplest nontrivial configuration necessary to produce nonclassicality in the 3-chain scenario which consists of 2-1-2 inputs (indicating the respective cardinalities $|X|$-$|Y|$-$|Z|$) and 2-2-2 outputs (respective cardinalities $|A|$-$|B|$-$|C|$). 

A correlation $p(a,b,c|x,z)$ will be classical in the 3-chain scenario iff it can be written as:

\begin{equation}
    p(a,b,c|x,y,z) = \int \int d\lambda_1 d\lambda_2 \mu(\lambda_1)\mu(\lambda_2) p_A(a|x,\lambda_1) p_B(b|y,\lambda_1,\lambda_2) p_C(c|z,\lambda_2).
    \label{eq:bilocalequation}
\end{equation}

\begin{figure}[h]
  \begin{subfigure}[b]{0.46\textwidth}
    \includegraphics[width=\textwidth, height=\textwidth]{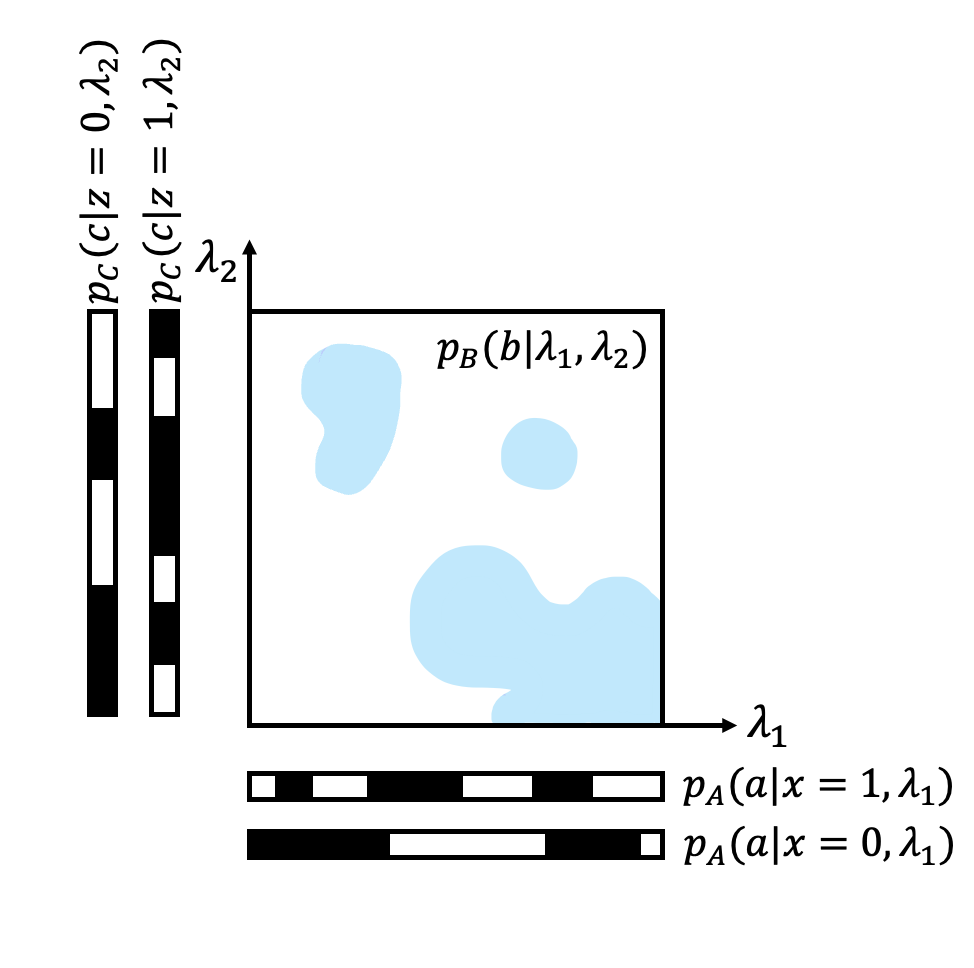}
    \caption{Before ordering.}
    \label{fig:f1}
  \end{subfigure}
  \hfill
  \begin{subfigure}[b]{0.46\textwidth}
    \includegraphics[width=\textwidth, height=\textwidth]{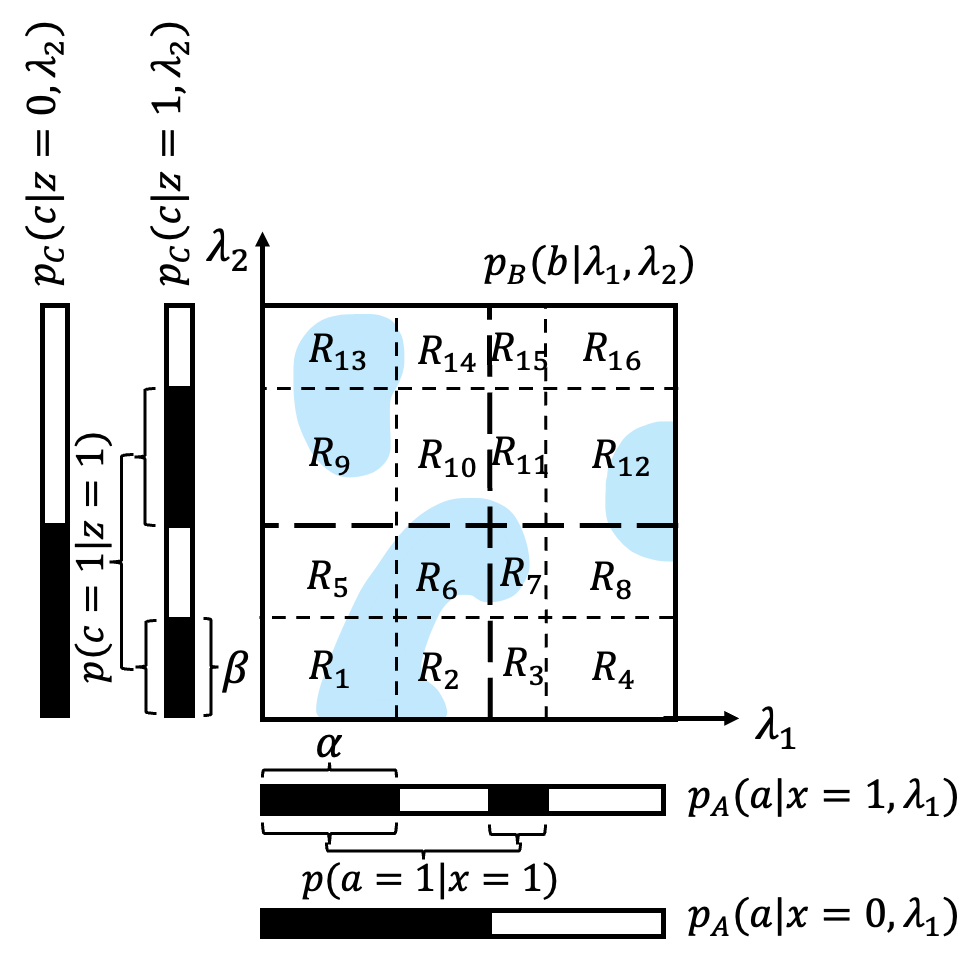}
    \caption{After ordering}
    \label{fig:f2}
  \end{subfigure}
  \caption{Pictorial representation of a classical correlation in the 3-chain scenario. The blue part corresponds to $p(a,b=1,c|x,z)$. The black regions in the bars correspond to $a=1$ in the $\lambda_1$-axis and $c=1$ in the $\lambda_2$-axis.}
  \label{fig:representation}
\end{figure}

The response functions in it can be assumed to be deterministic, as any randomness in them can be absorbed in $\mu(\lambda_1)$ and $\mu(\lambda_2)$. Furthermore, $\lambda_1$ and $\lambda_2$ can be taken as flat distributions from 0 to 1  (then, $\mu(\lambda_1) = 1 = \mu(\lambda_2)$), by rescaling the response function of each party. Hence, the equation~\eqref{eq:bilocalequation} becomes:

\begin{equation}
    p(a,b,c|x,y,z) = \int_0^1 \int_0^1 d\lambda_1 d\lambda_2 p_A(a|x,\lambda_1) p_B(b|\lambda_1,\lambda_2) p_C(c|z,\lambda_2).
\label{ec:initial}
\end{equation}
This means that correlations can be represented in a square as it is showed in Fig.~\ref{fig:f1}. Moreover, $\lambda_1$ and $\lambda_2$ are taken to be between 0 and 1 and their order can be permuted without loss of generality. Hence, we obtain Fig.~\ref{fig:f2}.

This way of visualising classical strategies divides the square into 16 rectangles ($R_{J}$ with $J=1,2,...,16$) each corresponding to one of the 16 possible deterministic strategies that Alice and Charlie decide upon receiving $\lambda_1$ and $\lambda_2$ respectively. In each rectangle, there can be a coloured part (which we will call $S_J$) and an uncoloured part that represent the probability of Bob outputting 1 or 0, respectively:
\begin{equation}
    S_J = p(b=1 \ \wedge\  \{\lambda_1,\lambda_2\}\in R_J).
\end{equation}
Moreover, there are two more variables that are unobservable, $\alpha$ and $\beta$, which correspond to another two degrees of freedom that can be visualized in Fig.~\ref{fig:f2}. Then, in order to fulfill the classicality condition, the variables $S_J$ must satisfy certain conditions.

First, since the variables $S_J$ corresponds to probabilities, they all have to fulfill positivity and have to be less than one,
\begin{equation}
    S_J\geq 0 \; \; \mbox{for} \;\; J=0,1,...,16.
    \label{eq:positivity}
\end{equation}
\begin{equation}
    S_J\leq 1 \; \; \mbox{for} \;\; J=0,1,...,16.
    \label{eq:lessone}
\end{equation}
Secondly, the variables $S_J$ have to be smaller than the rectangles they are contained in,
\begin{equation}
    S_J\leq R_J \;\; \mbox{for} \;\; J=0,1,...,16.
    \label{eq:rectangle}
\end{equation}
Eq.~\eqref{eq:rectangle} makes Eq.~\eqref{eq:lessone} redundant. All these conditions can be translated in terms of probabilities:
\begin{equation}
\begin{split}
S_1 &\leq \alpha \cdot \beta \\
S_2 &\leq \left[p(a=1,b,c|x=0,z=0)-\alpha\right]\cdot \beta\\
S_3 &\leq \left[p(a=1,b,c|x=1,z=0)-\alpha\right]\cdot \beta\\
S_4 &\leq \left[p(a=1,b,c|x=1,z=0)-p(a=1,b,c|x=0,z=0)+\alpha\right]\cdot \beta\\
S_5 &\leq \alpha \cdot  \left[p(a,b,c=1|x=0,z=0)-\beta\right]\\
S_6 &\leq \left[p(a=1,b,c|x=0,z=0)-\alpha\right]\cdot \left[p(a,b,c=1|x=0,z=0)-\beta\right]\\
S_7 &\leq \left[p(a=1,b,c|x=1,z=0)-\alpha\right]\cdot \left[p(a,b,c=1|x=0,z=0)-\beta\right]\\
S_8 &\leq \left[p(a=1,b,c|x=1,z=0)-p(a=1,b,c|x=0,z=0)+\alpha\right]\cdot \left[p(a,b,c=1|x=0,z=0)-\beta\right]\\
S_9 &\leq \alpha \cdot \left[p(a,b,c=1|x=0,z=1)-\beta\right]\\
S_{10} &\leq \left[p(a=1,b,c|x=0,z=0)-\alpha\right]\cdot \left[p(a,b,c=1|x=0,z=1)-\beta\right]\\
S_{11} &\leq \left[p(a=1,b,c|x=1,z=0)-\alpha\right]\cdot \left[p(a,b,c=1|x=0,z=1)-\beta\right]\\
S_{12} &\leq \left[p(a=1,b,c|x=1,z=0)-p(a=1,b,c|x=0,z=0)+\alpha\right]\cdot \left[p(a,b,c=1|x=0,z=1)-\beta\right]\\
S_{13} &\leq \alpha \cdot \left[p(a,b,c=1|x=0,z=1)-p(a,b,c=1|x=0,z=0)+\beta\right]\\
S_{14} &\leq \left[p(a=1,b,c|x=0,z=0)-\alpha\right]\cdot \left[p(a,b,c=1|x=0,z=1)-p(a,b,c=1|x=0,z=0)+\beta\right]\\
S_{15} &\leq \left[p(a=1,b,c|x=1,z=0)-\alpha\right]\cdot \left[p(a,b,c=1|x=0,z=1)-p(a,b,c=1|x=0,z=0)+\beta\right]\\
S_{16} &\leq \left[p(a=1,b,c|x=1,z=0)-p(a=1,b,c|x=0,z=0)+\alpha\right]\cdot \\
& \cdot \left[p(a,b,c=1|x=0,z=1)-p(a,b,c=1|x=0,z=0)+\beta\right].
\end{split}
\label{eq:group1}
\end{equation}
Additionally, some conditions on Bob's response function have to be satisfied. It is sufficient to look at $p(a,b=1,c|x,z)$ (the coloured part in Fig.~\ref{fig:representation}), since $p(a,b=0,c|x,z) = p(a|x)p(c|z) - p(a,b=1,c|x,z)$. These conditions are:
\begin{equation}
    \begin{split}
        &S_1 + S_2 + S_5 + S_6 = p(a=1,b=1,c=1|x=0,z=0)\\
        &S_3 + S_4 + S_7 + S_8 = p(a=0,b=1,c=1|x=0,z=0)\\
        &S_9 + S_{10} + S_{13} + S_{14} = p(a=1,b=1,c=0|x=0,z=0)\\
        &S_{11} + S_{12} + S_{15} + S_{16} = p(a=0,b=1,c=0|x=0,z=0)\\
        &S_1 + S_3 + S_5 + S_7 = p(a=1,b=1,c=1|x=1,z=0)\\
        &S_9 + S_{11} + S_{13} + S_{15} = p(a=1,b=1,c=0|x=1,z=0)\\
        &S_1 + S_2 + S_9 + S_{10} = p(a=1,b=1,c=1|x=0,z=1)\\
        &S_3 + S_4 + S_{11} + S_{12} = p(a=0,b=1,c=1|x=0,z=1)\\
        &S_1 + S_3 + S_9 + S_{11} = p(a=1,b=1,c=1|x=1,z=1).
    \end{split}
    \label{eq:group2}
\end{equation}
Finally, observing the pictorial representation (Fig.~\ref{fig:f2}) the last conditions for the unobserved variables are:
\begin{equation}
    \begin{split}
        \alpha &\geq 0 \\
        \alpha &\leq p(a=1|x=1)\\
        \alpha &\leq p(a=1|x=0)\\
        \alpha &\geq p(a=1|x=1)-p(a=0|x=0)\\
        \beta &\geq 0\\
        \beta &\leq p(c=1|z=1)\\
        \beta &\leq p(c=1|z=0)\\
        \beta &\geq p(c=1|z=1)-p(c=0|z=0),
    \end{split}
    \label{eq:group3}
\end{equation}
but the third, the fourth and the last two equations of \eqref{eq:group3} are redundant with the other constraints. Then, we have 18 variables and 45 constraints.\\

We can express our formulation more compactly (using marginal probability notation) as a standard optimization problem as follows:
\begin{align}
    \text{find} \quad & \alpha, \beta, S_n \quad n\in\{0,\ldots,15\} \\
    \text{s.t.}\quad & 0\leq S_{4i+j}\leq V_i \cdot H_j \quad i,j\in \{0,1,2,3\} \quad \text{where}:\\
    &  (V_i)_i = \begin{bmatrix}
        \beta \\
        p(c{=}1|z{=}0)-\beta \\
        p(c{=}1|z{=}1)-\beta \\
        p(c{=}1|z{=}1) - p(c{=}1|z{=}0) + \beta
        \end{bmatrix}, \quad (H_j)_j = \begin{bmatrix}
        \alpha \\
        p(a{=}1|x{=}0)-\alpha \\
        p(a{=}1|x{=}1)-\alpha \\
        p(a{=}1|x{=}1)-p(a{=}1|x{=}0)+\alpha
    \end{bmatrix}, \\
    & \max\left[0, p(a{=}1|x{=}1)-p(a{=}0|x{=}0)\right] \leq \alpha \leq \min\left[p(a{=}1|x{=}1)  ,  p(a{=}1|x{=}0)\right] \quad (H_j\geq 0),\\
    & \max\left[0, p(c{=}1|z{=}1)-p(c{=}0|z{=}0)\right] \leq \beta \leq \min\left[p(c{=}1|z{=}1),  p(c{=}1|z{=}0)\right] \quad (V_i\geq 0),\\
    & p(a,b{=}1,c|00) = \sum_{(i,j)\in \mathcal{I}^{00}_{ab}} S_{4i+j}, \quad \mathcal{I}^{00}_{ab} = \{ (2+n-2c,2+m-2a) \}_{n,m \in \{0,1\}}, \\
    & p(a{=}1,b{=}1,c|10) = \sum_{(i,j)\in \mathcal{I}^{10}_c} S_{4i+j},  \quad \mathcal{I}^{10}_c=\{(3-n-2c, 2m)\}_{n,m\in\{0,1\}} \\
    & p(a,b{=}1,c{=}1|01) = \sum_{(i,j)\in \mathcal{I}^{01}_a} S_{4i+j}, \quad \mathcal{I}^{01}_a=\{(2n+a,2+m-2a)\}_{n,m\in\{0,1\}}, \\
    & p(a{=}1,b{=}1,c{=}1|11) = \sum_{(i,j)\in \mathcal{I}^{11}} S_{4i+j}, \quad \mathcal{I}^{11} = \{ (2n,2m)\}_{n,m\in\{0,1\}}.
\end{align}

This formalism gives completely equivalent results to the one proposed in Ref.~\cite{branciard2012bilocal}, that uses a decomposition onto deterministic correlations, but the former reduces the number of variables. Codes for both formulations are available in Ref.~\cite{codes}. 

\clearpage
\section{Feasibility \mnn\ Linear Programs}
\label{feasibility_LPs}
\begin{figure}[tbh]
  \begin{subfigure}[b]{0.48\textwidth}
    \includegraphics[width=7cm]{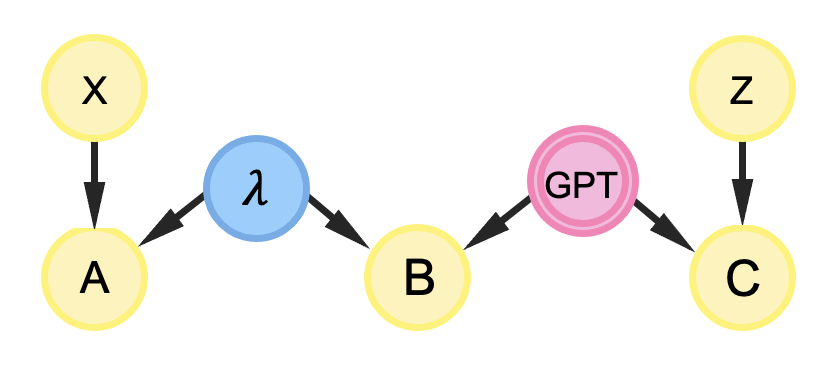}
    \caption{}
    \label{packed}
  \end{subfigure}
  \hfill
  \begin{subfigure}[b]{0.48\textwidth}
    \includegraphics[width=7cm]{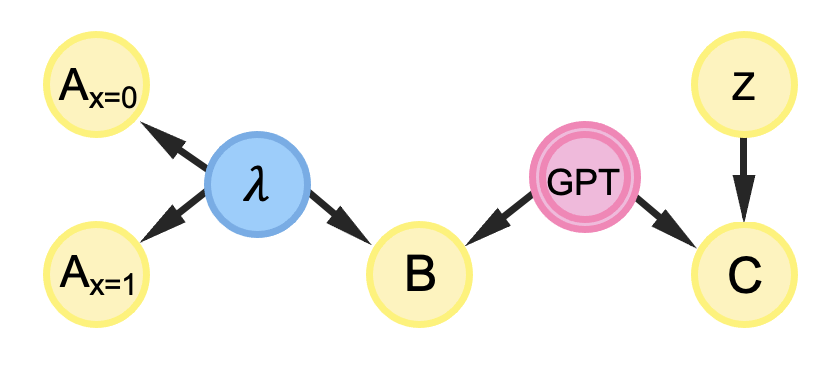}
    \caption{}
    \label{unpacked}
  \end{subfigure}
  \caption{Representation of the 3-chain scenario compatible with one classical source $\lambda$ and one OPT source. \ref{packed} illustrates the version before unpacking and \ref{unpacked}, after unpacking.}
  \label{fig:compatibility_oneclassical}
\end{figure}

We wanted to see whether a probability distribution is compatible with the the DAG shown in Fig. \ref{packed}. For a matter of convenience, we rewrite it in the unpacked version (see Fig.~\ref{unpacked}, and Ref.~\cite[Sec.5]{Navascues_2020} for the definition of unpacking). 

In order to determine if a correlation $p$ is or not compatible with the DAG previously mentioned, we need to establish whether there exists a probability distribution {\color{blue}$q(a_0, a_1, b, c| z)$} {\color{blue} on the DAG of Fig.~\ref{unpacked} }such that the following conditions are fulfilled:
\begin{itemize}
    \item No Signalling (NS).
    \item Independence.
    \item Compatibility with $p$.
\end{itemize}
This can be formulated as a standard feasibility problem as follows:
\begin{align}
    \text{find} \quad & q(a_0,a_1,b,c|z) \quad a_0,a_1,b,c,z \in \{ 0,1\} \\
    \text{s.t.}\quad & q(a_0,a_1,b,c|z) \geq 0 \quad \forall a_0,a_1,b,c,z,\\
    & \sum_{a_0,a_1,b,c} q(a_0,a_1,b,c|z) = 1 \quad \forall z, \\
    & q(a_0,a_1,b|z=0) = q(a_0,a_1,b|z=1) \quad \forall a_0,a_1,b\\
    & q(a_0,a_1,c|z) = q(a_0,a_1)q(c|z) \quad \forall a_0,a_1,c,z,\\
    & p(abc|x=0,z)= \sum_{a_1} q(a_0=a,a_1,b,c|z)\quad \forall a,b,c,z, \\
    & p(abc|x=1,z)= \sum_{a_0} q(a_0,a_1=a,b,c|z) \quad \forall a,b,c,z.
\end{align}
Note that $q(c|z)$ is constant, since Charlie's setting is not unpacked as he has access to an OPT source of correlations. Under these conditions, the program is linear and can be solved exactly using standard optimization methods.

We implemented this feasibility program using Gurobi~\cite{gurobi} and one can find more details in \cite{codes}.\\

For the case in which the source between Alice and Bob is OPT-compatible and the one between Bob and Charlie is classical, we followed an analogous procedure.

\clearpage
\section{Examples of \mnn correlations}
\label{examples_MNN}

This appendix provides detailed examples of minimal nonclassical correlations. We can define a ``post-selection'' box $p_{PS}$ given by the following correlations:
\begin{align}
    p_{PS}(a,b=0,c|x,z) &= p_{ps} \cdot p_{PR-box(V)}(a,c|x,z)\\
    p_{PS}(a,b=1,c|x,z) &= \frac{1}{4} - p_{ps} \cdot p_{PR-box(V)}(a,c|x,z),
\end{align}
where $p_{PR-box(V)}$ is given by
\begin{align}
    p_{PR-box(V)}(a,b|x,y) = (1 + V \cdot (-1)^{a + b + x \cdot y}) / 4.
\end{align}
The correlation coming from the post-selection box is parameterized by the visibility V and the probability of post-selection $p_{ps}$. Note that in the case of $V = 1/\sqrt{2}$ and $p_{ps} = 1/4$, it coincides with the entanglement-swapping correlation that we described on \ref{app:shadow_examples}.\\

We can also define a Bell local test $p_{local}$ in the following way. Bob and Charlie share a classical source $\lambda$, which can return 0 or 1. Bob's measurement is determined by $\lambda$ and Charlie ouputs $\lambda$ directly, ignoring $z$. Then, Charlie's outputs can be interpreted as Bob's inputs. Also, both Alice and Bob perform the same measurements: $\hat{A_0} = \hat{B_0} = (\sigma_x-\sigma_z)/\sqrt{2}$ and $\hat{A_1} = \hat{B_1} = (\sigma_x+\sigma_z)/\sqrt{2}$. Thus, producing a local correlation in the 3-chain scenario. Explicitly, the produced correlations are as follows:
\begin{align}
    p(a,b,c|x,z) = p(c|z)\cdot p(a,b|x,c)\quad
    \mbox{where}\quad p(c|z) = \frac{1}{2}\\ \mbox{and}\quad p(a,b|x,c) = \begin{dcases*} 1/4\quad \mbox{if}\quad x \oplus c = 1 \\
    1/2 \quad\mbox{if}\quad x\oplus c = 0 \; \mbox{and} \; a\oplus b =0\\
    0 \quad\mbox{if}\quad x\oplus c = 0 \; \mbox{and} \; a\oplus b =1
    \end{dcases*}
\end{align}

The cases of minimal network nonclassicality appear when we convexly mix the correlations previously mentioned (Eq. \ref{eq:convex}) for a certain range of $\mu$ given $V\in(1/\sqrt{2},1)$ and $p_{ps} = 1/4$.

\begin{align}
    p(a,b,c|x,z) = \mu \cdot p_{Post-selection}(V, p_{ps}) + (1-\mu)\cdot p_{local}.
    \label{eq:convex}
\end{align}
In particular, we know that the case of $V=1/\sqrt{2}$ and $p_{ps} = 1/4$ is quantum, because this correlation can be obtained in a protocol where Alice always performs the same measurements on the same states, ensuring that the correlation obtained by means of the convex combination is also quantum. This case is \mnn\ for $\mu \in (0.455, 0.705)$.

In addition, we also found post-quantum correlations certified via quantum inflation utilizing the Inflation library \cite{boghiu2023inflation}. These are given by the same convex combination when $V=1$ and $p_{ps}=1/4$ and they are \mnn\ for $\mu \in (0, 0.5)$.

Finally, it is noteworthy that a modified version of the protocol proposed in \cite{lauand2024quantum} is also \mnn. In this case, the state shared by Alice and Bob and the one by Bob and Charlie is a maximally entangled state, concretely $\ket{\phi^+} =(\ket{00}+\ket{11})/2$. The measurements performed by Alice and Charlie are $\hat{A}_0 = \hat{C}_0 = \sigma_x$ and $\hat{A}_1 = \hat{C}_1 = \sigma_z$ for their inputs $x,z=0,1$, respectively. Bob has two possible outcomes and the respective measurements are $\hat{B}(b=0) = \ket{\psi_{\theta}}\bra{\psi_{\theta}}$ and $\hat{B}(b=1) = \mathbb{1} - \ket{\psi_{\theta}}\bra{\psi_{\theta}}$, where $\ket{\psi_{\theta}} = \sin{\theta}\ket{01} + \cos{\theta}\ket{10}$. This protocol is \mnn for $\theta \in (0,\pi/2)\setminus\theta/4$.

\clearpage
\section{Observational causal order for the triangle scenario}
\label{app: triangle_causal_hierarchy}

\begin{figure}[h!]
\centering
\includegraphics[width=10cm]{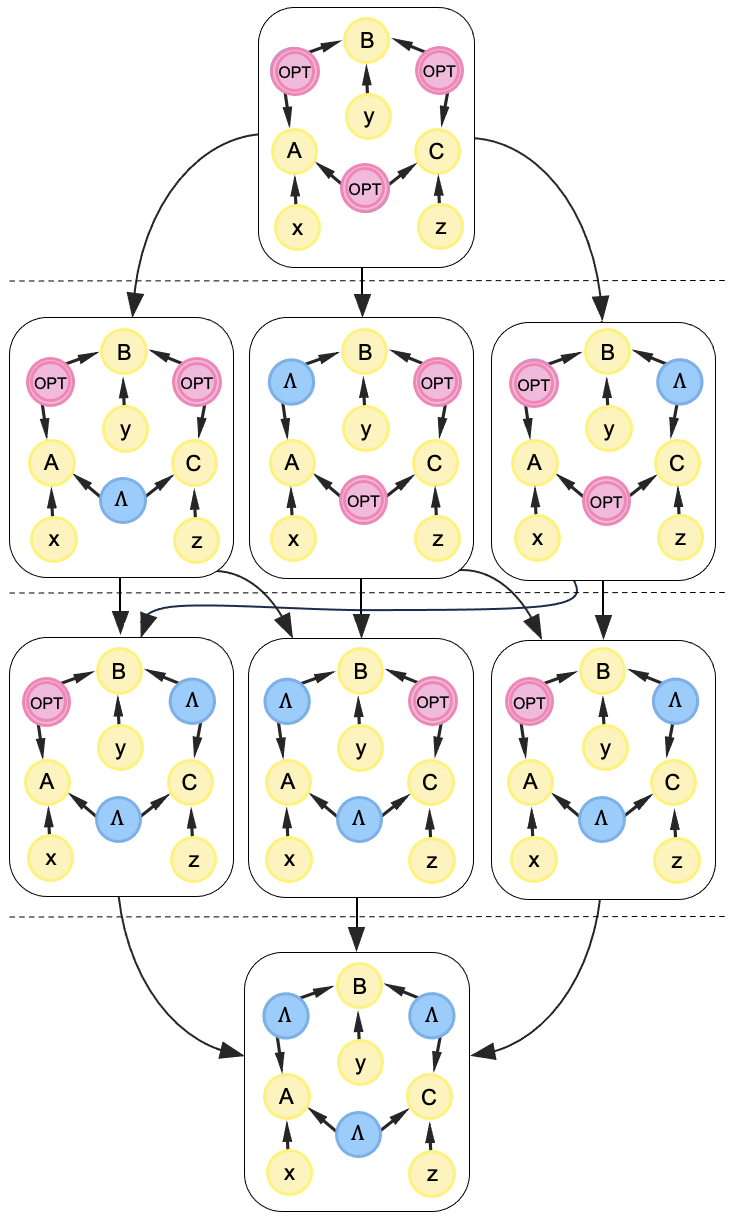}
\caption{Representation of the observational causal order for the triangle network. Outside arrows represent dominance (i.e. the set of correlations compatible with one scenario are included in the set of correlations compatible with the other). $\Lambda$ represents a classical source, while $OPT$, an OPT one.}
\label{fig:causal_order_triangle}
\end{figure}

\end{document}